\documentclass[aps,prd,preprintnumbers,nofootinbib]{revtex4}
\usepackage[utf8]{inputenc}
\usepackage{empheq}
\usepackage{bm}
\usepackage[normalem]{ulem}
\usepackage{latexsym}
\usepackage{color}
\usepackage{dcolumn}
\usepackage{amsmath,amsfonts,amssymb}
\usepackage{graphicx,epsfig}
\usepackage{psfrag}
\usepackage{amsthm}
\usepackage[pdftex]{hyperref}
\usepackage{amsmath}
\usepackage[toc,title,page]{appendix}
\usepackage{natbib}
\usepackage{amsfonts}
\def\be{\begin{eqnarray}}
\def\ee{\end{eqnarray}}

\begin{document}

\title{The Squeezed OTOC and Cosmology}
\author{S. Shajidul Haque}\email{shajid.haque@uct.ac.za}
\affiliation{High Energy Physics, Cosmology \& Astrophysics Theory Group \\and The Laboratory for Quantum Gravity \& Strings,\\
Department of Mathematics and Applied Mathematics,\\
University of Cape Town, South Africa} 
\author{Bret Underwood}\email{bret.underwood@plu.edu}
\affiliation{Department of Physics,\\
Pacific Lutheran University,\\
Tacoma, WA 98447}

\date{\today}

\vspace{-2cm}
\begin{abstract}

Exponential growth in the out-of-time-order correlator (OTOC) is an important potential signature of quantum chaos.
The OTOC is quite simple to calculate for 
squeezed states, whose applications are frequently found in quantum optics and cosmology.
We find that the OTOC for a generic highly squeezed quantum state is exponentially large, suggesting that highly squeezed states are ``primed'' for quantum chaos.
A quantum generalization of the classical symplectic phase space matrix can be used to extract the quantum Lyapunov spectrum, and we find this better captures the exponential growth of squeezed states for all squeezing angles compared to any single OTOC.
By describing cosmological perturbations in the squeezed state language, we are able to apply our calculations of the OTOC to arbitrary expanding and contracting backgrounds 
with fixed equation of state.
We find that only expanding de Sitter backgrounds support an exponentially growing OTOC at late times, with a putative Lyapunov exponent consistent with other calculations.
While the late-time behavior of the OTOC for other cosmological backgrounds appears to change depending on the equation of state, we find that the quantum Lyapunov spectrum shows some universal behavior: the OTOC grows proportional to the scale factor for perturbation wavelengths larger than the cosmological Hubble horizon.

\end{abstract} 

\maketitle
\section{Introduction}

Unlike classical systems, characterizing the nature of chaos in quantum mechanics and quantum many body systems can be challenging. 
Classical chaotic systems are characterised by a hypersensitivity to perturbations in initial conditions through time evolution, so that nearby trajectories diverge exponentially. 
This behavior can be captured by the Poisson bracket between the position and momentum variables
\begin{eqnarray}
   \left\{q(t),p(0)\right\}^2 = \left(\frac{\partial q(t)}{\partial q(0)}\right)^{2}\sim \sum_{n}c_{n}e^{2\lambda_{n}t}\,,
   \label{ClassicalChaos} 
\end{eqnarray}
where the $\lambda_n$ are known as Lyapunov exponents.
In quantum mechanics, the analogue of (\ref{ClassicalChaos}) is the \emph{unequal time commutator} $[\hat q(t),\hat p(0)]$, which becomes the Poisson bracket $\sim i\hbar \{q(t),p(0)\}$ in the semiclassical limit.
Exponential growth of this commutator can potentially serve as an indication of quantum chaos, reflecting the growing overlap between these operators.
However, the unequal-time commutator is in general an operator, not a $c$-number, so a more useful measure is the double unequal-time commutator, also known as the \emph{out-of-time-order correlator} (OTOC)
\begin{equation}
 {\cal C}(t) \equiv - \langle \left[ \hat q(t), \hat p(0) \right]^2 \rangle_{\beta}\, , \
 \label{outoftimedefinition}
\end{equation}
where angle brackets $\langle \cdots \rangle_{\beta}$ denote an average over some set of states, typically taken as a thermal average at temperature $T = 1/\beta$.
More generally, one can consider the OTOC of any two Hermitian operators
\be
{\cal C}_{W,V}(t) = - \langle \left[\hat W(t),\hat V(0)\right]^2\rangle_\beta\, .
\ee
A related quantity, originally considered in the context of superconductivity and becoming increasingly important in high energy and condensed matter physics, is the out-of-time-order four-point function
\cite{Kitaev2015,Larkin1969,Maldacena:2015waa}, 
\be
 F(t) \equiv \langle \hat W(t) \hat V(0) \hat W(t) \hat V(0) \rangle_{\beta} \,.
 \label{4ptoutoftimedefinition}
\ee
The two are commonly related through ${\cal C}_{V,W}(t) \sim 2\left(\langle \hat W\hat W\rangle\langle\hat V\hat V\rangle - \mathrm{Re}\left(F(t)\right)\right)$ at late times, so that their information content is approximately the same and are usually referred to interchangeably.
In this work we will refer to the squared commutator (\ref{outoftimedefinition}) as the OTOC, and will not focus on the closely related four-point function (\ref{4ptoutoftimedefinition}).
Further, as will be seen, the commutators considered in this work are $c$-numbers, 
so our results will be completely independent of the averaging procedure.

In a quantum chaotic system, by analogy with the classical case we expect the OTOC to exhibit exponential growth ${\cal C}(t) \sim e^{2\lambda t}$ at early times, 
characterized by a a \emph{quantum Lyapunov exponent} $\lambda$, up to some later time, after which it saturates.
However, the single OTOC fails to capture chaotic behavior for single-particle quantum chaotic systems, such as the well-known stadium billiards model \cite{Hashimoto:2017oit,Rozenbaum_2017,Rozenbaum_2019}, or chaotic lattice systems, such as spin chains \cite{Gharibyan_2019}.
Indeed, the OTOC is not the only possible probe of quantum chaos in a system. A web of diagnostics of quantum chaos \cite{bhattacharyya2019web,Kudler-Flam:2019kxq} -- including the OTOC, quantum circuit complexity, and others -- can be useful for studying quantum chaos from different angles and perspectives.
One useful extension of the OTOC is the quantum Lyapunov spectrum \cite{QuantLyapunovSpectrum}, a generalization of the classical matrix of phase space deviations.
We will compare our results of the OTOC with the quantum Lyapunov spectrum, finding the latter to be more consistent in describing the late-time dynamics.

While the OTOC has been used to investigate the dynamics and potential quantum chaotic behavior of several specific quantum systems,
in this paper we will calculate the OTOC for the general class of states known as squeezed states.
Squeezed states, in which the uncertainty of the position and momentum operators
is squeezed in some phase space direction while their product is still minimized, are characterized by a squeezing parameter $r$ and a squeezing angle $\phi$.
Squeezed states arise in diverse contexts, from models of quantum optics, gravitational wave detection, or descriptions of cosmological perturbations, and can serve as a useful toy model for many interesting systems.
Since squeezed states are freely interacting, the OTOC can be calculated in closed form and is independent of the OTOC averaging.
We find that the OTOC for highly squeezed states is exponentially large, implying that highly squeezed states are ``primed'' for quantum chaos.
Indeed, applying our formalism of squeezed states to the well-studied inverted harmonic oscillator, we find exponential growth in the OTOC, consistent with other work \cite{me3,Bhattacharyya:2020art}.
It is well known that the inverted harmonic oscillator is not strictly chaotic; since it is unbounded from below, growth in the OTOC thus more properly reflects its unstable nature. However, it serves as an exactly solvable toy model for studying potential diagnostics of quantum chaos in quantum field theories \cite{Blume_Kohout_2003,Morita_2019,bueno2019complexity,Yan_2020,Betzios_2016,Hegde_2019,Hashimoto_2020}. 
We will see that the matrix of unequal-time commutators used in constructing the quantum Lyapunov spectrum for squeezed states has a symplectic structure, as in the classical case, and can be used to extract general OTOC behavior of the system that is not sensitive to the squeezing angle itself.

Contemporary interest 
in quantum information theory has uncovered the OTOC as an important ingredient in the AdS/CFT correspondence, and the OTOC plays an important role in the study of the scrambling of information by black holes \cite{Hayden_2007,Sekino_2008}.
In this paper, however, we are interested in the application of the OTOC to cosmology, in order to gain potentially new and complementary insights about the behaviour of cosmological perturbations and models of the early Universe such as inflation.
Using quantum circuit complexity as a diagnostic of quantum chaos, \cite{us,us2} demonstrated that two-mode squeezed state cosmological perturbations in an expanding, accelerating universe may be chaotic in nature, with a Lyapunov exponent bounded from above\footnote{See also \cite{ChoudhuryOTOC,ChoudhuryComplexity} for related work on the OTOC and circuit complexity in the context of cosmology.}.
Given the difficulty of understanding the nature of quantum chaos by a single diagnostic, in this paper we will examine the behavior of the OTOC of cosmological perturbations in expanding and contracting cosmological backgrounds with fixed equation of state.

We begin the paper in Section \ref{sec:Inverted} with an analysis of the OTOC of squeezed states in quantum mechanics to set up the tools and techniques of our approach, including a generalization to the squared matrix of OTOCs. 
As a concrete example, we consider the inverted harmonic oscillator in the language of squeezed states, and calculate the corresponding OTOC.
Cosmological perturbations in Friedmann-Lema\^{i}tre-Robertson-Walker (FLRW)  backgrounds are naturally described in Fourier space in the language of two-mode squeezed states \cite{Grishchuk,Albrecht,Martin1,Martin2} (see also \cite{BellCMB1,BellCMB2} for an application of squeezed states to Bell CMB experiments), so, in Section \ref{sec:Squeezed} we develop the formalism for the OTOC of two-mode squeezed states for continuous Fourier modes, mirroring the analysis of Section \ref{sec:Inverted}.
We apply the squeezed state formalism to cosmological perturbations and their OTOC for expanding and contracting backgrounds with fixed equation of state in Section \ref{sec:CosmoOTOC}.
One cosmological background of particular interest, because of its similarities with black hole backgrounds and its relevance to early- and late-time expansion of the Universe, is de Sitter space.
Interestingly, we find that expanding de Sitter space is the only cosmological background that exhibits late-time exponential growth of the OTOC.
In Appendix \ref{sec:dSOTOC}, we calculate the OTOC of Fourier modes for de Sitter space using the known exact solutions for the mode functions, recovering the result of Section \ref{sec:CosmoOTOC}, and extend this analysis to the OTOC of de Sitter cosmological perturbations in position-space. Note that the OTOC of 3d de Sitter space has also been studied recently by \cite{Shiu} with different techniques, and is in excellent agreement with our results here.
Finally, we conclude with some discussion of our results and their potential interpretation in Section \ref{sec:Discussion}.


\section{Squeezed States and the Out-of-Time-Ordered Correlator}
\label{sec:Inverted}

We will begin by introducing the main techniques and concepts used throughout the rest of the paper in the context of quantum mechanics, and apply our formalism to the inverted harmonic oscillator, where the language of squeezed states arises naturally.
Our fundamental building blocks will the annihilation and creation operators $\hat a$ and $\hat a^\dagger$, which act in the usual way on the vacuum $\hat a |0\rangle = 0$ and eigenstates $|n\rangle$ of the number operator.
The single-mode squeezed vacuum state $|r,\phi\rangle = \hat {\cal S}(r,\phi)|0\rangle$, given in terms of the (time-dependent) squeezing parameter $r$ and squeezing angle $\phi$, is generated by the action of the squeezing operator $\hat {\mathcal S}(r,\phi)$ on the vacuum, which can be written as
\be
\hat {\mathcal S}(r,\phi) \equiv {\rm exp}\ \left[\frac{r}{2} \left(e^{-2i\phi} \hat a^2 - e^{2i\phi} {\hat a^{\dagger 2}}\right)\right]\, .
\ee
We will be interested in unitary evolution $\hat {\mathcal U}(\eta,\eta_0)$ (where will will use the variables $\eta,\eta_0$ for time unless otherwise noted) that can be parameterized in the factorized form \cite{Grishchuk,Albrecht}
\be
\hat {\mathcal U}(\eta,\eta_0) = \hat {\mathcal S}(r,\phi)\ \hat {\mathcal R}(\theta)\, ,
\ee
for potentially time-dependent $r(\eta),\phi(\eta),\theta(\eta)$ where $\hat {\mathcal R}(\theta)$ is the rotation operator written in terms of the time-dependent rotation parameter $\theta$
\be
\hat {\mathcal R}(\theta) = {\rm exp}\ \left[-i\theta \hat a^\dagger \hat a\right]\, .
\ee
In the Heisenberg picture, $\hat {\mathcal U}$ acting on the creation and annihilation operators generates the time evolution
\be
\hat a(\eta) \equiv \hat {\mathcal U}(\eta,\eta_0)^\dagger\ \hat a\ \hat {\mathcal U}(\eta,\eta_0) = \cosh r\ e^{-i\theta}\ \hat a^{(0)} - \sinh r\ e^{+i(\theta+2\phi)}\ \hat a^{(0) \dagger}\, .
\label{SqueezedA1}
\ee
Note that the squeezing and rotation evolution operators preserve the equal-time commutation relations between the creation and annihilation operators, so that if $[\hat a^{(0)},\hat a^{(0)\dagger}] = 1$ at some initial time, then $[\hat a(\eta),\hat a(\eta)^\dagger] = 1$ at later times as well.

We define position and momentum operators for some fundamental frequency $k$ in the usual way
\be
\hat q = \frac{1}{\sqrt{2k}} \left(\hat a^\dagger + \hat a\right), \hspace{.4in} \hat p = i \sqrt{\frac{k}{2}} \left(\hat a^\dagger - \hat a\right)\, .
\label{InvertCreation}
\ee
Using (\ref{SqueezedA1}) and (\ref{InvertCreation}) we can rewrite the unequal-time commutator\footnote{See also \cite{Ando} for other related work on unequal-time correlators.} between $\hat q$ and $\hat p$ as
\be
[\hat q(\eta),\hat p(\eta_0)] = \frac{i}{2} \left[\left((\cosh r\ e^{-i\theta} - \sinh r\ e^{-i(\theta+2\phi)}\right)\left(\cosh r_0\ e^{i\theta_0} + \sinh r_0\ e^{i(\theta_0+2\phi_0)}\right) + {\rm c.c.}\right]\, .
\label{UTC1General}
\ee
Taking the initial state to be ``unsqueezed'' $r_0 \rightarrow 0$, (\ref{UTC1General}) becomes
\be
[\hat q(\eta),\hat p(\eta_0)] = i \left[ \cosh r\,\cos(\theta-\theta_0) - \sinh r\, \cos(\theta-\theta_0+2\phi)\right]\, .
\label{InvertedUTC1}
\ee
Note that in the limit $\eta\rightarrow \eta_0$, this agrees with the equal-time commutator above.
In order to understand the significance of (\ref{InvertedUTC1}), let us take some specific values of the parameters $r,\phi,\theta$. In particular, for a squeezing angle that is a multiple of $\pi$, (\ref{InvertedUTC1}) is exponentially small in the squeezing parameter
\be
[\hat q(\eta),\hat p(\eta_0)] = i e^{-r}\ \cos(\theta-\theta_0)\, ,\hspace{.4in} \mbox{for $\phi = n\pi$}
\label{UTC1}
\ee
while for a squeezing angle that is an odd multiple of $\pi/2$ it is exponentially large in the squeezing parameter
\be
[\hat q(\eta),\hat p(\eta_0)] = i e^{r}\ \cos(\theta-\theta_0)\, , \hspace{.4in} \mbox{for $\phi = \frac{2n+1}{2}\pi$}
\ee
More generally, for any squeezing angle that is not an integer multiple of $\pi$, the unequal time commutator becomes exponentially large at large squeezing
\be
[\hat q(\eta),\hat p(\eta_0)] \sim i e^{r}\, ,\hspace{.4in} \mbox{for $\phi \neq n\pi$}
\ee

As discussed in the Introduction, one common measure of quantum chaos is the {\bf out-of-time-order correlator} (OTOC)
\be
{\mathcal C}(\eta) \equiv \langle-\left[\hat q(\eta),\hat p(\eta_0)\right]^2\rangle_\beta\, ,
\label{OTOCDef1}
\ee
where the averaging $\langle \cdots\rangle_\beta$ over states is typically a 
thermal expectation value for temperature $\beta = 1/T$
\be
\langle \hat {\mathcal O}\rangle_\beta \equiv \frac{{\rm Tr}\,\left( e^{-\beta \hat H} \hat {\mathcal O}\right)}{{\rm Tr}\, \left(e^{-\beta \hat H}\right)}\, .
\ee
Note that for systems controlled by a quadratic Hamiltonian such as squeezed states, the thermal expectation value of the position-momentum commutator 
(\ref{InvertedUTC1}) is a $c$-number, so that the OTOC (\ref{OTOCDef1}) 
is \emph{independent} of the averaging procedure 
\be
{\mathcal C}(\eta) = -\left[\hat q(\eta),\hat p(\eta_0)\right]^2\, .
\label{OTOCDef2}
\ee
Since the OTOC (\ref{OTOCDef2}) is simply proportional to the square of the unequal-time commutator (\ref{UTC1General}), we will occasionally refer to the commutator and its square interchangeably as the OTOC throughout the text. 
We can define a {\bf \it quantum Lyapunov exponent} $\lambda$ from ${\mathcal C}(\eta)$ when there is exponential growth in time of the OTOC ${\mathcal C}(\eta) \sim e^{2\lambda \eta}$.
Taking a highly squeezed state (with $\phi \neq n\pi$), the OTOC is  exponentially large
\be
{\mathcal C}(\eta) \sim e^{2r}\, .
\ee
Certainly, even though the OTOC is exponentially large, such a highly squeezed state itself is not necessarily in a state of quantum chaos, since it is the exponential dependence on \emph{time} that defines a system with quantum chaos.
However, because of the exponential dependence, even any linear time-dependent deviation from constant squeezing
$r\approx r_0 + \lambda \eta$ 
will exhibit quantum chaos with a Lyapunov exponent $\lambda$.
Thus, in this respect a highly squeezed state is ``primed'' for quantum chaos.

Interestingly, a highly squeezed state with $\phi = n\pi$ with this same linear growth in squeezing exhibits what appears to be the opposite of quantum chaos: an exponential \emph{decrease} in the OTOC, as in (\ref{UTC1}).
This \emph{quantum attractor}-like behavior for a specific squeezing angle in the quantum phase space is an artifact of restricting ourselves to the single unequal-time commutator (\ref{InvertedUTC1}), and motivates a generalization of the OTOC.
For classical chaos, the sensitivity to initial conditions can be investigated by studying the symplectic matrix
\be
M_{ij}(\eta) \equiv \frac{\delta z_i(\eta)}{\delta z_j(\eta_0)},
\label{ClassicalM}
\ee
where the $z_i$ run over the canonical coordinates $x,p$.
The Lyapunov exponents can then be extracted from eigenvalues of the squared matrix
\be
L_{ij} = \left[M^\dagger(\eta) M(\eta)\right]_{ij}\, .
\label{ClassicalL}
\ee
A natural extension of (\ref{ClassicalM}),(\ref{ClassicalL})
to quantum mechanics is the matrix of unequal-time commutators \cite{QuantLyapunovSpectrum}
\be
\hat {\mathcal M}_{ij}(\eta) \equiv -i[\hat Q_i(\eta),\hat Q_j(\eta_0)]\, ,
\label{QuantumM}
\ee
and its averaged square
\be
{\cal L}_{ij} \equiv \left\langle \left[\hat {\cal M}^\dagger(\eta) \hat {\cal M}(\eta)\right]_{ij} \right\rangle_\beta\, ,
\label{QuantumL}
\ee
where we have normalized $\hat Q_1 = \sqrt{k}\,\hat q$ and $\hat Q_2 = \hat p/\sqrt{k}$.
The relevant unequal time commutators for the elements of (\ref{QuantumM}), under the assumption of an unsqueezed initial state, together with (\ref{InvertedUTC1}) are
\be
\left[\hat p(\eta),\hat q(\eta_0)\right] &=& -i \left[ \cosh r\,\cos(\theta-\theta_0) + \sinh r\, \cos(\theta-\theta_0+2\phi)\right]\, ; \label{FieldMomentumUTC}\\
\left[\hat q(\eta), \hat q(\eta_0)\right] &=& \frac{i}{k} \left[\cosh r\, \sin(\theta-\theta_0) - \sinh r\, \sin(\theta-\theta_0+2\phi)\right]\, ; \label{FieldFieldUTC}\\
\left[\hat p(\eta),\hat p(\eta_0)\right] &=& i k \left[\cosh r\, \sin(\theta-\theta_0) + \sinh r\, \sin(\theta-\theta_0+2\phi)\right]\, . \label{MomentumMomentumUTC}
\ee
Note that since the unequal-time commutators (\ref{InvertedUTC1}),(\ref{FieldMomentumUTC})-(\ref{MomentumMomentumUTC}) are $c$-numbers for squeezed states, the matrix elements $\hat {\cal M}_{ij}$ are also $c$-numbers, and so the squared matrix ${\cal L}$ is independent of the thermal averaging.
It is straightforward to check that the matrix $\hat {\mathcal M}$ has unit determinant for squeezed states; since it is also a 2 x 2 matrix, it is therefore necessarily {\it symplectic}, similarly as its classical counterpart.
This formulation of the squared matrix of out-of-time-order correlators is preferable to focusing on the single OTOC constructed using $[\hat q(\eta),\hat p(\eta_0)]$, because it avoids a reliance in one particular direction in the phase space of canonical coordinates.

The eigenvalues $\alpha_{\pm}$ for ${\cal L}$ can easily be calculated, though their general form is not particularly enlightening.
Recall that the OTOC (\ref{UTC1}) for $\phi = n \pi$ was exponentially small for large squeezing, in contrast to the behavior at a generic angle.
The eigenvalues of ${\cal L}$ for $\phi = n \pi$,
\be
\alpha_{\pm} = \cosh(2r)\sin^2(\theta-\theta_0) \pm \sqrt{\cos(2(\theta-\theta_0)) + \cosh^2(2r) \sin^4(\theta-\theta_0)}\, ,
\ee
instead illustrate that there is indeed a large eigenvalue $\sim e^{2r}$ for large squeezing.
More generally, for an arbitrary angle $\phi$, in the large squeezing limit $r \gg 1$ the dominant eigenvalue of ${\cal L}$ scales as $\alpha_- \sim e^{2r}$, 
so that the presence of an exponentially large eigenvalue is a generic consequence of a highly squeezed state.
It is tempting, then, to identify the generalized behavior of the growth of the OTOC as the dominant eigenvalue of ${\cal L}$.

Up to this point we have been speaking in general terms about squeezed states without reference to any specific system.
For concreteness, we can apply the general formalism of squeezed states developed above in the context of the quantum inverted harmonic oscillator. The Hamiltonian is a harmonic potential that is unbounded from below
\be
\hat H = \frac{1}{2} \hat p^2 - \frac{1}{2} k^2 \hat q^2 = -\frac{k}{2}\left(\hat a^2 + \hat a^{\dagger 2}\right)\, ,
\label{InvertedH}
\ee
where we took a unit mass $m = 1$, and we rewrote the Hamiltonian in terms of the raising and lowering operators.
The time evolution of the squeezing and rotation parameters $r,\phi,\theta$ are determined by the Heisenberg equation of motion
\be
\frac{d}{d\eta} \hat a = i\left[\hat H,\hat a\right]\, .
\ee
For the Hamiltonian (\ref{InvertedH}), this leads to the equations of motion for the squeezing parameters
\be
\dot r &=& k \sin( 2\phi)\, ; \nonumber \\
\dot \phi &=& k \coth(2r) \cos(2\phi)\, ; \\
\dot \theta &=& 0\, . \label{InvertedEOM}
\ee
It is easy to see that these equations have a solution in which the squeezing grows with time along a constant squeezing angle with constant rotation angle
\be
r(\eta) = k \eta, \hspace{.4in} \phi(\eta) = \pi/4\, ,\hspace{.4in} \theta(\eta) = \theta_0\, .
\label{InvertedSoln}
\ee
Inserting (\ref{InvertedSoln}) into the position-momentum unequal time commutator (\ref{InvertedUTC1}), we obtain at late times $k\eta \gg 1$
\be
[\hat q(\eta),\hat p(\eta_0)] = i \cosh (k\eta) \sim \frac{i}{2} e^{k\eta}\, .
\label{InvertedUTC2}
\ee
with corresponding OTOC from (\ref{OTOCDef2})
\be
{\mathcal C}(\eta) = \cosh^2(k\eta)\sim \frac{1}{4} e^{2k\eta} \sim e^{2\lambda \eta}\, .
\label{InvertedOTOC}
\ee
Since (\ref{InvertedUTC2}) approaches an exponential at late times, we can identify a {\it quantum Lyapunov exponent} for the inverted harmonic oscillator $\lambda \approx k$.
We can also insert these solutions into the matrix $\hat {\cal M}$ (\ref{QuantumM}) of unequal-time commutators, and extract the Lyapunov exponent from the eigenvalues of ${\cal L}$ (\ref{QuantumL}).
With $\theta = \theta_0$ and $\phi = \pi/4$, the 
matrix $\hat {\cal M}$ takes the simple form
\be
\hat{\cal M} = \begin{pmatrix}
-\sinh r & \cosh r \cr
-\cosh r & \sinh r \cr
\end{pmatrix}\, ,
\ee
while the eigenvalues $\alpha_{\pm}$ of ${\cal L}$ take the simple form
\be
\alpha_{\pm} = \left\{-1,\cosh(2r)\right\} \sim \left\{-1, e^{2k\eta}\right\}\, ,
\ee
for $r = k\eta \gg 1$.
We see that the dominant eigenvalue $e^{2k\eta}$ captures the exponential growth we see in 
the single OTOC (\ref{InvertedOTOC}), as expected, identifying the Lyapunov exponent $\lambda \sim k$ as before.

\section{Squeezed States in Quantum Field Theory and the OTOC}
\label{sec:Squeezed}
The squeezed state formalism described in the previous section applies equally well in quantum field theory, with minor modifications. In particular, we will decompose a free real quantum scalar field $\hat v(\vec{x},\eta)$ into creation and annihilation operators
\be
\hat v(\vec{x},\eta)
&=& \int \frac{d^3k}{(2\pi)^3} e^{i\vec{k}\cdot \vec{x}} \hat v_{\vec{k}}(\eta) = \frac{1}{\sqrt{2k}} \int \frac{d^3k}{(2\pi)^3} \left(\hat a_{\vec{k}}(\eta) e^{i\vec{k}\cdot \vec{x}} + \hat a^\dagger_{\vec{k}}(\eta) e^{-i\vec{k}\cdot \vec{x}} \right)\, ,
\ee
where 
$[\hat a_{\vec{k}},\hat a^\dagger_{\vec{k}'}] = (2\pi)^3 \delta^3(\vec{k}-\vec{k}')$ and $\hat a_{\vec{k}} |0\rangle = 0$ as usual.
The momentum-preserving {\it two-mode squeezing operator} can be written as 
\be
\hat {\mathcal S}_{\vec{k}}(r_k,\phi_k) \equiv {\rm exp}\ \left[\frac{r_k(\eta)}{2} \left(e^{-2i\phi_k(\eta)} \hat a_{\vec{k}} \hat a_{-\vec{k}} - e^{2i\phi_k(\eta)} \hat a_{-\vec{k}}^\dagger \hat a_{\vec{k}}^\dagger\right)\right]\, .
\ee
The unitary evolution for a general quadratic, momentum-preserving Hamiltonian is generated by
\be
\hat {\mathcal U}_{\vec{k}} = \hat{\mathcal S}_{\vec{k}}(r_k,\phi_k) \hat{\mathcal R}_{\vec{k}}(\theta_k)\, ,
\ee
where as before $\hat{\mathcal R}_{\vec{k}}$ is the two-mode rotation operator
\be
\hat{\mathcal R}_{\vec{k}}(\theta_k) \equiv {\rm exp}\ \left[-i\theta_k(\eta) (\hat a_{\vec{k}} \hat a_{\vec{k}}^\dagger + \hat a_{-\vec{k}}^\dagger \hat a_{-\vec{k}})\right]\, ,
\ee
written in terms of the rotation angle parameter $\theta_k(\eta)$ and $\hat{\mathcal S}_{\vec{k}}$.
As before, the evolution of the unitary operator $\hat {\mathcal U}_{\vec{k}}$ acting on the creation and annihilation operators generates the time evolution
\be
\hat a_{\vec{k}}(\eta) &=& \hat {\mathcal U}_{\vec{k}}^\dagger \hat a_{\vec{k}}(0) \hat {\mathcal U}_{\vec{k}} = \cosh r_k\ e^{-i\theta_k}\hat a_{\vec{k}}(0) - \sinh r_k\ e^{i(\theta+2\phi_k)}\ \hat a_{-\vec{k}}(0)\, ; \label{FTaSqueeze}\\
\hat a^\dagger_{\vec{k}}(\eta) &=& \hat {\mathcal U}_{\vec{k}}^\dagger \hat a^\dagger_{\vec{k}}(0) \hat {\mathcal U}_{\vec{k}} = -\sinh r_k\ e^{-i(\theta_k + 2 \phi_k)}\ \hat a_{-\vec{k}}(0) + \cosh r_k\ e^{i\theta_k}\ \hat a^\dagger_{\vec{k}}(0)\,.\label{FTadagSqueeze}
\ee

We will write the Fourier modes of the field $\hat v_{\vec{k}}(\eta)$ and its momentum $\hat \pi_{\vec{k}}(\eta)$ in terms of the raising and lowering operators
\be
\hat v_{\vec{k}} &=& \frac{1}{\sqrt{2k}} \left[\hat a_{\vec{k}}(\eta) + \hat a^\dagger_{-\vec{k}}(\eta)\right]\, ; \\
\hat \pi_{\vec{k}} &=& -i\sqrt{\frac{k}{2}} \left[\hat a_{-\vec{k}}(\eta) - \hat a^\dagger_{\vec{k}}(\eta)\right]\, .
\ee
Using (\ref{FTaSqueeze}),(\ref{FTadagSqueeze})
for the raising and lowering operators in terms of the squeezing parameters, we can rewrite the field and its momentum as
\be
\hat v_{\vec{k}}(\eta) &=& \frac{1}{\sqrt{2k}} \left(\cosh r_k\ e^{-i\theta_k} - \sinh r_k\ e^{-i(\theta_k + 2\phi_k)}\right) \hat a_{\vec{k}}(0) + \frac{1}{\sqrt{2k}}\left(\cosh r_k\ e^{i\theta_k} - \sinh r_k\ e^{i(\theta_k + 2\phi_k)}\right) \hat a_{-\vec{k}}^\dagger(0)\, ; \\
\hat \pi_{\vec{k}}(\eta) &=& -i \sqrt{\frac{k}{2}} \left(\cosh r_k\ e^{-i\theta_k} + \sinh r_k e^{-i(\theta_k+2\phi_k)}\right) \hat a_{-\vec{k}}(0) + i\sqrt{\frac{k}{2}} \left(\cosh r_k\ e^{i\theta_k} + \sinh r_k e^{i(\theta_k+2\phi_k)}\right) \hat a_{\vec{k}}^\dagger(0)\, .
\ee

From these expressions it is relatively straightforward to calculate the momentum-space unequal-time commutator
\be
[\hat v_{\vec{k}}(\eta),\hat \pi_{\vec{k}'}(\eta_0)] 
&=& i (2\pi)^3 \delta^3(\vec{k}-\vec{k}') f_k(\eta,\eta_0)\, ,
\label{FTUTC1}
\ee
where we defined
\be
f_k(\eta,\eta_0) = \frac{1}{2}\left[\left((\cosh r_k\ e^{-i\theta_k} - \sinh r_k\ e^{-i(\theta_k+2\phi_k)}\right)\left(\cosh r_0\ e^{i\theta_0} + \sinh r_0\ e^{i(\theta_0+2\phi_0)}\right) + {\rm c.c.}\right]\, ,
\label{fGeneral}
\ee
as the amplitude of the unequal-time commutator, with the delta function and other factors of $(2\pi)$ and $i$ stripped off, and a zero in the subscript means the parameter is evaluated at $\eta = \eta_0$, e.g.~$r_0 = r_k(\eta_0)$.
Notice once again that taking equal times $\eta \rightarrow \eta_0$ leads to $f_k(\eta_0,\eta_0) \rightarrow 1$, as it should for the equal time commutator.

As with the quantum mechanics case, we will define a Fourier mode OTOC as the double commutator
\be
{\mathcal C}_{\vec{k}}(\eta) \equiv - \langle [\hat v_{\vec{k}_1}(\eta),\hat \pi_{\vec{k}_1'}(\eta_0)][\hat v_{\vec{k}_2}(\eta),\hat \pi_{\vec{k}_2'}(\eta_0)]\rangle_{\beta}\, ,
\ee
where we used two copies of the unequal time commutator with different momenta to avoid delta function divergences. Once again, the unequal time commutator is a $c$-number, so the averaging is trivial and the OTOC simply becomes the square of the commutator
\be
{\mathcal C}_{\vec{k}}(\eta) = - [\hat v_{\vec{k}_1}(\eta),\hat \pi_{\vec{k}_1'}(\eta_0)][\hat v_{\vec{k}_2}(\eta),\hat \pi_{\vec{k}_2'}(\eta_0)]\, .
\label{FTOTOC}
\ee
The OTOC (\ref{FTOTOC}) is proportional to the square of the amplitude of the unequal-time commutator (\ref{fGeneral}), and some factors of $(2\pi)$ and delta-functions enforcing momentum conservation
\be
{\mathcal C}_{\vec{k}}(\eta) = (2\pi)^6 f_{k_1}(\eta,\eta_0) f_{k_2}(\eta,\eta_0) \delta^3\left(\vec{k}_1 - \vec{k}_1'\right)\delta^3\left(\vec{k}_2-\vec{k}_2'\right)\, .
\ee
Ignoring the delta-functions and other factors, then, we have
\be
{\mathcal C}_{\vec{k}}(\eta) \sim f_{k_1}(\eta,\eta_0) f_{k_2}(\eta,\eta_0)\, ,
\label{FTOTOC2}
\ee
so that the OTOC and the amplitude $f_k(\eta,\eta_0)$ have roughly the same behavior for $\vec{k}_1 \sim \vec{k}_2 \sim \vec{k}$.
Because of this, we will mostly focus our analysis on the amplitude function $f_k(\eta,\eta_0)$, which we will treat as synonymous with the OTOC.

As in Section \ref{sec:Inverted}, we can construct the matrix
$\hat {\mathcal M}$ of canonical unequal-time commutators (\ref{QuantumM}), with elements
\be
\hat {\mathcal M}_{ij} \equiv -i\left[\hat Q_i(\eta),\hat Q_j(\eta_0)\right]\, ,
\label{CosmoM}
\ee
for $\hat Q_1 = \sqrt{k}\ \hat v_{\vec{k}}$ and $\hat Q_2 = \hat \pi_{\vec{k}'}/\sqrt{k}$.
We will similarly define the squared matrix ${\cal L}$ as in (\ref{QuantumL})
Correspondingly, it is straightforward to calculate the analogs of (\ref{FieldMomentumUTC})-(\ref{MomentumMomentumUTC}) for the field theory case; the results are, unsurprisingly, identical to the quantum mechanics case except for additional factors of $(2\pi)^3 \delta^3(\vec{k}-\vec{k}')$. 
In particular, we will write the unequal-time commutators as
\be
\left[\hat \pi_{\vec{k}'}(\eta),\hat v_{\vec{k}}(\eta_0)\right] &=& -i (2\pi)^3 g_k(\eta,\eta_0) \delta^3\left(\vec{k}-\vec{k}'\right)\, ; \\
\left[\hat v_{\vec{k}}(\eta),\hat v_{\vec{k}'}(\eta_0)\right] &=& i (2\pi)^3 h_k(\eta,\eta_0) \delta^3\left(\vec{k}-\vec{k}'\right)\, ; \\
\left[\hat \pi_{\vec{k}}(\eta),\hat \pi_{\vec{k}'}(\eta_0)\right] &=& i (2\pi)^3 j_k(\eta,\eta_0) \delta^3\left(\vec{k}-\vec{k}'\right)\, ,
\ee
where
\be
g_k(\eta,\eta_0) &=& \frac{1}{2}\left[\left((\cosh r_k\ e^{i\theta_k} + \sinh r_k\ e^{i(\theta_k+2\phi_k)}\right)\left(\cosh r_0\ e^{-i\theta_0} - \sinh r_0\ e^{-i(\theta_0+2\phi_0)}\right) + {\rm c.c.}\right]\, ; \label{gGeneral} \\
h_k(\eta,\eta_0) &=& \frac{1}{2}\left[\left((\cosh r_k\ e^{i\theta_k} - \sinh r_k\ e^{i(\theta_k+2\phi_k)}\right)\left(\cosh r_0\ e^{-i\theta_0} - \sinh r_0\ e^{-i(\theta_0+2\phi_0)}\right) - {\rm c.c.}\right]\, ;\label{hGeneral} \\
j_k(\eta,\eta_0) &=& \frac{1}{2}\left[\left((\cosh r_k\ e^{-i\theta_k} + \sinh r_k\ e^{-i(\theta_k+2\phi_k)}\right)\left(\cosh r_0\ e^{i\theta_0} + \sinh r_0\ e^{i(\theta_0+2\phi_0)}\right) - {\rm c.c.}\right]\, . \label{jGeneral}
\ee
Together with $f_k(\eta,\eta_0)$ in (\ref{fGeneral}), these functions represent the amplitudes of the unequal-time commutators.

For an unsqueezed initial state $r_0 \rightarrow 0$, the functions $f_k(\eta,\eta_0) - j_k(\eta,\eta_0)$ become
\be
f_k(\eta,\eta_0) &\approx & \cosh r_k\ \cos(\theta_k-\theta_0) - \sinh r_k\ \cos(\theta_k-\theta_0+2\phi_k)\, ; \label{fUnsqueezed}\\
g_k(\eta,\eta_0) &\approx & \cosh r_k\ \cos(\theta_k-\theta_0) + \sinh r_k\ \cos(\theta_k-\theta_0+2\phi_k)\, ; \label{gUnsqueezed}\\
h_k(\eta,\eta_0) &\approx & \cosh r_k\ \sin(\theta_k-\theta_0) - \sinh r_k\ \sin(\theta_k-\theta_0+2\phi_k)\, ;\label{hUnsqueezed}\\
j_k(\eta,\eta_0) &\approx & \cosh r_k\ \sin(\theta_k-\theta_0) + \sinh r_k\ \sin(\theta_k-\theta_0+2\phi_k)\, . \label{jUnsqueezed}
\ee
Again, the matrix $\hat {\cal M}$ then has determinant one, and 
we will use the dominant eigenvalue of ${\cal L}$ to identify the Lyapunov exponent, as discussed in Section \ref{sec:Inverted}.

We will find in Section \ref{sec:CosmoOTOC} that the assumption of an unsqueezed initial state does not cover all cases of interest.
In particular, we will also be interested in the case where the initial state is highly squeezed $r_0 \gg 1$, with a squeezing angle $\phi_0 \approx -\pi/2$ and vanishing rotation angle $\theta_0 \approx 0$.
In this limit, the amplitude functions (\ref{fGeneral}),(\ref{gGeneral})-(\ref{jGeneral}) become
\be
f_k(\eta,\eta_0) &\approx & e^{-r_0}\left(\cosh r_k\ \cos(\theta_k) - \sinh r_k\ \cos(\theta_k+2\phi_k)\right)\, ; \\
g_k(\eta,\eta_0) &\approx & e^{-r_0}\left(\cosh r_k\ \cos(\theta_k) + \sinh r_k\ \cos(\theta_k+2\phi_k)\right)\, ; \\
h_k(\eta,\eta_0) &\approx & e^{-r_0}\left(\cosh r_k\ \sin(\theta_k) - \sinh r_k\ \sin(\theta_k+2\phi_k)\right)\, ; \\
j_k(\eta,\eta_0) &\approx & e^{-r_0}\left(\cosh r_k\ \sin(\theta_k) + \sinh r_k\ \sin(\theta_k+2\phi_k)\right)\, .
\ee
Clearly, these are similar to the unsqueezed initial state, up to a common multiplicative factor of $e^{-r_0}$. 
The eigenvalues of the corresponding ${\cal L}$ will therefore be identical to those of the unsqueezed initial state, with appropriate choices of $r_k,\theta_k,\phi_k$.
Finally, we will also be interested in the initial conditions $r_0 \gg 1$ with vanishing angles $\phi_0,\theta_0 \approx 0$. The amplitude functions will once again mirror the structure of (\ref{fUnsqueezed}) - (\ref{jUnsqueezed}), up to a common multiplicative factor of $e^{+r_0}$ this time.

In the next subsection, we will apply this language of two-mode squeezed states to study the behavior of the OTOC (\ref{FTOTOC}) for cosmological perturbations in an expanding background.

\section{Cosmological Perturbations and the OTOC}
\label{sec:CosmoOTOC}

We begin our analysis by reviewing the description of cosmological perturbations \cite{Mukhanov} as two-mode squeezed states, following \cite{us} (see also \cite{Grishchuk,Albrecht,Martin1,Martin2}).
We take as our background a spatially flat Friedmann-Lemaitre-Robertson-Walker (FLRW) metric
\be
ds^2 = -dt^2 + a(t)^2 d\vec{x}^2 = a(\eta)^2 \left(-d\eta^2+d\vec{x}^2\right)\, ,
\label{FLRW}
\ee
where we have introduced the conformal time $\eta$.
For simplicity we will consider our matter content to consist of a fundamental scalar field $\varphi$ with a canonical kinetic term and potential $V(\varphi)$. On the background (\ref{FLRW}) the scalar field is homogeneous and time-dependent $\varphi_0(t)$; the scalar field potential and kinetic energy source the evolution of the scalar field $a(\eta)$.

On this background we will consider fluctuations of a scalar field $\varphi(x) = \varphi_0(t) + \delta\varphi(t,x)$ with the metric
\be
ds^2 = a(\eta)^2 \left[-(1+2\psi(x,\eta))d\eta^2 + (1-2\psi(x,\eta)) d\vec{x}^2\right]\, .
\ee
In terms of the curvature perturbation ${\mathcal R} = \psi+\frac{H}{\dot \varphi_0} \delta \varphi$, where a dot denotes a derivative with respect to cosmic time $t$, and the Hubble parameter $H = \dot a/a$, the action takes the following simple form \cite{Mukhanov}
\be
S = \frac{M_{pl}^2}{2}\int dt\, d^3x\, a^3 \frac{\dot \varphi_0^2}{H^2} \left[\dot {\mathcal R}^2 - \frac{1}{a^2} \left(\partial_i {\mathcal R}\right)^2\right]\, .
\label{CosmoAction0}
\ee
By using the Mukhanov variable $v \equiv z {\mathcal R}$
where $z \equiv M_{pl}\, a\, \sqrt{2\epsilon}$, with $\epsilon = -\dot H/H^2 = 1-{\mathcal H}'/{\mathcal H}^2$, the action can be transformed into the following form:
\be
S = \frac{1}{2} \int d\eta\, d^3x \left[v'^2 - (\partial_i v)^2 + \left(\frac{z'}{z}\right)^2 v^2 - 2 \frac{z'}{z} v' v\right]\, ,
\label{CosmoAction}
\ee
where the prime denotes a derivative with respect to conformal time and ${\mathcal H} = a'/a$.
This action represents perturbations of a scalar field coupled to an external time-varying source\footnote{A virtually identical-looking expression can also be derived for tensor perturbations with the replacement $z'/z \rightarrow a'/a$, and our results will hold for these types of perturbations as well.}.
We promote the perturbation to a quantum field and expand into Fourier modes $\hat v_{\vec{k}}$, with the action
\be
S = \frac{1}{2} \int d\eta\ d^3k\left[\hat v_{\vec{k}}' \hat v_{-\vec{k}}' - \left(k^2-\left(\frac{z'}{z}\right)^2\right) \hat v_{\vec{k}} \hat v_{-\vec{k}} - 2 \frac{z'}{z} \hat v_{\vec{k}} \hat v_{-\vec{k}}'\right]\, .
\ee
The corresponding Hamiltonian can be written in terms of the raising and lowering operators as 
\be
\hat H = \int d^3k\, \hat {\mathcal H}_{\vec{k}} = \int d^3k \left[k\left(\hat a_{\vec{k}} \hat a_{\vec{k}}^\dagger + \hat a_{-\vec{k}}^\dagger \hat a_{-\vec{k}}\right) 
    - i \frac{z'}{z} \left(\hat a_{\vec{k}} \hat a_{-\vec{k}} - \hat a_{\vec{k}}^\dagger \hat a_{-\vec{k}}^\dagger\right)\right]\, .
\label{CosmoH}
\ee
The first term in (\ref{CosmoH}) represents the usual free-particle Hamiltonian, while the second term describes the interaction between the quantum perturbation and the expanding background.
Notice that this last term is similar in form to the Hamiltonian for the inverted harmonic oscillator from the last section, and indeed we will see that when the last term in the Hamiltonian dominates $z'/z \gg k$ the dynamics will resemble that of an inverted harmonic oscillator.
The time evolution of the Fourier modes generated by ${\cal H}_{\vec{k}}$ is naturally described by the product of the two-mode squeezing ${\cal S}_{\vec{k}}(r_k,\phi_k)$ and rotation ${\cal R}_{\vec{k}}(\theta_k)$ operators, as outlined in Section \ref{sec:Squeezed}.

The Heisenberg equation of motion $d\hat a_{\vec{k}}/d\eta =i \left[\hat {\mathcal H}_k,\hat a_{\vec{k}}\right]$, together with the squeezed form of the creation and annihilation operators (\ref{FTaSqueeze}),(\ref{FTadagSqueeze})
leads to the equations of motion for the squeezing parameters
\be
\frac{dr_k}{d\eta} &=& -\frac{z'}{z} \cos (2\phi_k)\, ;~  \\
\frac{d\phi_k}{d\eta}&=& k + \frac{z'}{z} \coth(2r_k) \sin (2\phi_k)\, ;  \\
\frac{d\theta_k}{d\eta} &=& k - \frac{z'}{z} \tanh(r_k) \sin(2 \phi_k)\, . \label{squeezeEOMeta}
\ee
It is often more convenient to rewrite these equations of motion with the scale factor $a(\eta)$ as the independent parameter, leading to
\begin{eqnarray}
\frac{dr_k}{da} &=& -\frac{1}{a}\,\cos(2\phi_k) \, ;
\label{squeezeEOM1} \\
\frac{d\phi_k}{da} &=& \frac{k}{a{\cal H}}
+ \frac{1}{a}\, \coth(2r_k) \sin(2\phi_k)\, ;
\label{squeezeEOM2}\\
\frac{d\theta_k}{da} &=&-\frac{k}{a{\cal H}} - \frac{1}{a} \tanh r_k \sin(2\phi_k)\, . \label{squeezeEOM3}
\end{eqnarray}

Solutions to (\ref{squeezeEOM1})-(\ref{squeezeEOM3}) are known \cite{us2}; in the rest of this section, we will use these known solutions to compute the unequal-time commutator (\ref{fGeneral}) and the OTOC (\ref{FTOTOC}).

\subsection{Expanding Backgrounds}

The scale factor in conformal time for expanding FLRW backgrounds (\ref{FLRW}) with a fixed equation of state $p = w\rho$ takes the form\footnote{As discussed in \cite{us2}, the marginal case $w = -1/3$ is uninteresting for our purposes, and we will not consider it further here.}
\begin{equation}
    a(\eta) = \left(\frac{\eta_0}{\eta}\right)^\beta = 
    \begin{cases}
    \left(\frac{\eta_0}{\eta}\right)^\beta, & -\infty < \eta < 0, ~ \eta_0 < 0 ~ \mbox{for accelerating backgrounds $\beta > 0$ ($w < -1/3$)} \\
    \left(\frac{\eta}{\eta_0}\right)^{|\beta|}, & 0 < \eta < \infty, ~ \eta_0 > 0 ~ \mbox{for decelerating backgrounds $\beta < 0$ ($w > -1/3$)}\, ,
    \end{cases}
    \label{ExpandingScaleFactor}
\end{equation}
where the parameter $\beta \equiv -\frac{2}{1+3w}$ is positive for accelerating solutions and negative for decelerating solutions.
The dimensionful parameter $\eta_0$ controls the (cosmic-time) Hubble expansion rate $H_0$ at the time $\eta = \eta_0$ as $|\eta_0| =  |\beta| H_0$. For our purposes, $|\eta_0|$ will show up in the combination $k|\eta_0|$, which controls when modes are sub-horizon 
or super-horizon.

For all accelerating backgrounds, $w < -1/3$, modes begin at sufficiently early times $a \ll (k|\eta_0|)^\beta$ inside the horizon with small squeezing $r_k \ll 1$, with the approximate solution
\be
r_k(a) &\approx& \frac{\beta}{2k|\eta_0|} a^{1/\beta} \ll 1\, ; \label{AccelSmallScaleSolnSqueeze}\\
\phi_k(a) &\approx& -\frac{\pi}{4} - \frac{1}{4k|\eta_0|} a^{1/\beta}\, ; \label{AccelSmallScaleSolnAngle}\\
\theta_k(a) &\approx & -\frac{k |\eta_0|}{a^{1/\beta}} \gg 1\, .\label{AccelSmallScaleSolnTheta}
\ee
Taking the initial state to be ``unsqueezed'' $r_0 \rightarrow 0$, the amplitude of the unequal-time commutator (\ref{fGeneral}) becomes
\be
f_k(\eta,\eta_0) = \cosh r_k \cos(\theta_k - \theta_0) - \sinh r_k \cos(\theta_k-\theta_0+2\phi_k)\, .
\label{fCosmo}
\ee
For times $\eta > \eta_0$, but still early enough that modes remain within the horizon, (\ref{fCosmo}) oscillates with approximately unit amplitude
\be
f_k(\eta,\eta_0) \approx \cos\left(\theta_k-\theta_0 \right) \sim {\mathcal O}(1)\, .
\ee
Thus, the OTOC for sufficiently early times when the modes are deep within the horizon is ${\cal C}_{\vec{k}}\sim {\cal O}(1)$.
At late times, once the modes exit the horizon $a \gg (k|\eta_0|)^\beta$,
the squeezing parameter grows, while the squeezing and rotation angles ``freeze out'',
with the leading-order solution \cite{us2}
\be
r_k(a) & \approx & \ln \left(a/k|\eta_0|\right)\, ; \\
\phi_k(a) &\approx & -\frac{\pi}{2}\, ; \\
\theta_k(a) &\approx & \frac{k |\eta_0|}{2\beta-1} \frac{1}{a^{1/\beta}} \ll 1\, .
\ee
At late times, then, the unequal-time commutator becomes dominated by the large squeezing
\be
f_k(\eta,\eta_0) \approx e^{\ln (a/k |\eta_0|)}\cos(\theta_0) = \frac{a(\eta)}{k|\eta_0|}\cos (\theta_0) \, .
\ee
The corresponding OTOC is then the square of this
\be
{\cal C}_k(\eta) \sim f_k(\eta,\eta_0)^2 \approx \left(\frac{a(\eta)}{k|\eta_0|}\right)^2\cos^2(\theta_0)\, .
\label{OTOCExpandAccel}
\ee

Up to now, we have been working with conformal time $\eta$, because the Hamiltonian and equations of motion take on a particularly simple form in this coordinate system.
In order to interpret our result (\ref{OTOCExpandAccel}) physically, however, it is more natural to evaluate it in terms of the cosmic time $t$, defined in (\ref{FLRW}), 
because this is the time observed by an observer comoving with the expanding fluid.
The qualitative behavior of the time-dependence of the accelerating solutions can be subdivided according to the null energy condition. When applied to a homogeneous and isotropic perfect fluid described by a pressure and energy density, the null energy condition $T_{\mu\nu} N^\mu N^\nu \geq 0$ becomes a condition on the equation of state $w \geq -1$.
Accelerating cosmological backgrounds satisfying the null energy condition $-1 < w < -1/3$ are generally realizable with known matter sources, including canonically normalized scalar fields; the corresponding scale factor takes a power-law form $a(t) \sim (t/t_0)^{2/(3(1+w))}$, where $a(t_0) = 1$. Cosmological backgrounds that saturate the null energy condition $w = -1$, including de Sitter space, experience exponential expansion $a(t) \sim e^{H (t-t_0)}$.
Finally, for cosmological backgrounds that violate the null energy condition $w < -1$, which may be more difficult to realize using simple matter fields without introducing additional pathologies, the scale factor diverges in finite time $a(t)\sim (t_e-t)^{2/(3(1+w))}$ at time $t_e$.
The corresponding time-dependence of the OTOC (\ref{OTOCExpandAccel}) follows that of the scale factor
\be
{\cal C}_k(t) \sim \begin{cases}
t^{\frac{4}{3(1+w))}} & \mbox{ for } -1 < w < -1/3\  (1 < \beta < \infty)  \cr
\left(\frac{H}{k}\right)^2 e^{2H t} & \mbox{ for } w = -1\  (\beta = 1) \cr
(t_e-t)^{-\frac{4}{3(1+w))}} & \mbox{ for }  w < -1\  (0 < \beta < 1) 
\end{cases}
\label{ExpandAccelOTOCLate}
\ee
As discussed in the introduction, we will schematically describe the growth of the OTOC ${\cal C}_k(t)$ between the Fourier mode and its momentum in terms of the semiclassical limit of the corresponding Poisson bracket.
For a chaotic system we expect the OTOC to grow exponentially with time;
interestingly, we find that \emph{only de Sitter solutions ($w=-1$) display strictly exponential growth of the OTOC},
\be
{\cal C}_k(t) \sim {\cal A} e^{2\lambda t} = \left(\frac{H_{dS}}{k}\right)^2\ e^{2 H t}\, ,
\label{dSLyapunov}
\ee
where $\lambda = H_{dS}$ is a Lyapunov exponent.
Putting this together with the early-time behavior, then, we see that the OTOC for a de Sitter background initially oscillates with unit amplitude, then grows exponentially with cosmic time after horizon crossing, as seen in Figure \ref{fig:dSOTOC}.
Following \cite{Maldacena:2015waa}, the timescale $t_* \sim (2\lambda)^{-1} \log (1/{\cal A})$ for the OTOC of a given mode with wavelength $k$ to become ${\cal O}(1)$ is
\be
t_*(k) \sim \frac{1}{2H_{dS}} \log \left(\frac{k}{H_{dS}}\right)^2 \sim H_{dS}^{-1} \log \frac{k}{H_{dS}}\, .
\label{tScramble}
\ee
{By taking the wavelength to be bounded above by a cutoff $k < \Lambda$, an estimate for an upper bound on this timescale for the entire system could be $t_* \leq H^{-1} \ln(\Lambda/H_{dS})$.
For large $N$ quantum field theories, it is more appropriate to describe the behavior of the system in terms of a four-point function $F(t) = \langle \hat W(t) \hat V(0) \hat W(t) \hat V(0)\rangle_\beta \approx 1 -  e^{\lambda t}/N$ for intermediate times longer than the typical thermalization timescale $t_d$ \cite{Maldacena:2015waa}.
A corresponding scrambling time is then defined as the timescale in which the four-point function decays, $t_{scr} \approx \lambda^{-1} \log N$ \cite{Maldacena:2015waa}; large $N$ ensures a hierarchy between the thermalization and scrambling times.
In our case, it is difficult to directly calculate the thermalization timescale $t_d$ in our formalism from the thermal two-point function, so it is not clear if such a hierarchy naturally exists. Nevertheless, a naive estimate would give the thermalization timescale to be set by the Gibbons-Hawking de Sitter temperature scale \cite{Gibbons:1977mu} $t_d \sim T^{-1} \sim H_{dS}^{-1} \sim \lambda^{-1}$.
Given a relationship between the OTOC and the four-point function in this limit of the form ${\cal C}(t) \approx 2 - 2 F(t) \approx 2 e^{\lambda t}/N$, the two timescales $t_*$ and $t_{scr}$ then might play analogous roles.

\begin{figure}[t]
\centering \includegraphics[width=.5\textwidth]{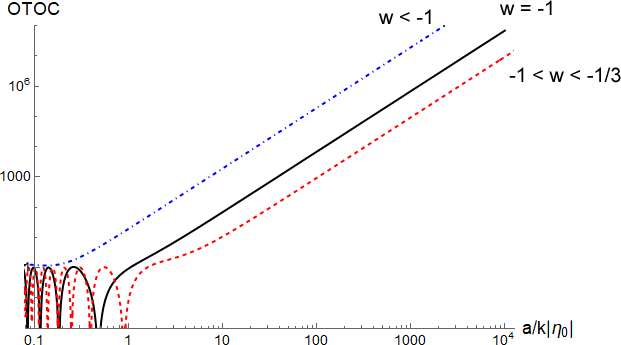}\hspace{.2in}\includegraphics[width=.46\textwidth]{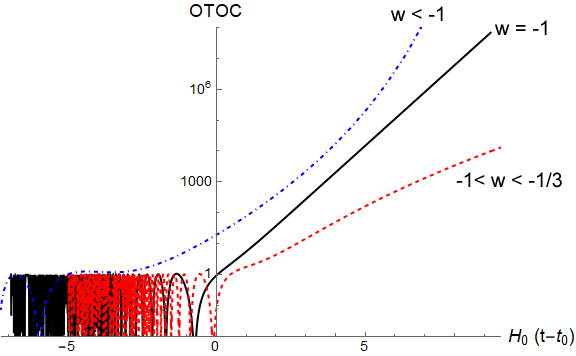}
\caption{The numerical solution to the squeezing equations of motion (\ref{squeezeEOM1})-(\ref{squeezeEOM3}) show the behavior of the OTOC ${\cal C}_k$ as a function of scale factor (left) and as a function of cosmic time $t$ (right) (written here in terms of $H_0(t-t_0)$, where $a(t_0) = 1$ and $H(t_0) = H_0$) for an expanding accelerating cosmological background depends on the equation of state. For all equations of state the OTOC is oscillatory with ${\cal O}(1)$ amplitude at early times, then grows as $a^2$ when the mode exits the horizon, following (\ref{ExpandAccelOTOCLate}). For accelerating equations of state between $-1 < w < -1/3$ (dashed red, $w=-0.9$), the OTOC grows as a power of $t$, slower than exponential. For $w = -1$ (solid black), the OTOC grows exponentially, implying potential quantum chaos for de Sitter expansion. For $w < -1$ (dash-dotted blue, $w=-1.1$), the growth of the OTOC is ``super-chaotic'', growing faster than exponential and diverging in finite time.}
\label{fig:dSOTOC}
\end{figure}

The putative Lyapunov exponent from (\ref{dSLyapunov}) also saturates the conjecture \cite{Maldacena:2015waa} that the Lyapunov exponent of the OTOC is bounded above by the temperature $\lambda \leq 2\pi T$, taking $T \sim H/2\pi$ \cite{Gibbons:1977mu}.
However, we again note that for our case the correspondence is less clear: the conjectured bound relies on a parametric separation of scales between the typical scattering, or thermalization, time $t_{d}$ and the scrambling time $t_{scr}$, which in our case might not be present.
Since the OTOC calculated using the unequal-time commutator (\ref{FTOTOC}) is done in Fourier space and is a $c$-number, it is independent of the thermal averaging procedure as also discussed earlier around Eqs.(\ref{OTOCDef2}) and (\ref{FTOTOC}).
Further, it is difficult using our formalism to determine the thermalization time, and so it is unclear that the assumptions and reasoning that went into \cite{Maldacena:2015waa} apply in our case.
It is interesting to compare our results for the OTOC (\ref{dSLyapunov}) and $t_*$ (\ref{tScramble}) to those of \cite{Shiu,Geng:2020kxh}.
While the Lyapunov exponents agree, up to perhaps a factor of two, (\ref{dSLyapunov}) does not manifestly vanish in the limit in which gravity is decoupled for a fixed de Sitter scale $M_{pl}/H_{dS} \rightarrow \infty$, while the OTOC of \cite{Shiu} does manifestly vanish in this limit.
Similarly, $t_*$ does not contain an explicit factor of $M_{pl}/H_{dS}$ in this decoupling limit.
Note, however, that this is an artifact of the transformation to Mukhanov variables below (\ref{CosmoAction0}), which  contains an explicit factor of $M_{pl}$. In the $M_{pl}/H_{dS} \rightarrow \infty$ limit, the curvature perturbations ${\cal R}$ described by the action (\ref{CosmoAction0}) do indeed decouple.
Nevertheless, (\ref{dSLyapunov}) does not lead to the same parametric control of the separation of time scales as \cite{Shiu} in this limit.
Note also that (\ref{dSLyapunov}) is evaluated in cosmic time with comoving coordinates, while the analysis of \cite{Shiu} uses static time coordinates, making it difficult to make a direct correspondence between these results.
Finally, note that while the Mukhanov variable $v$ grows at late times for de Sitter, the corresponding curvature perturbation ${\cal R}$ is constant on these superhorizon scales, so we do not expect the linearized theory to break down due to gravitational backreaction.

Considering accelerating solutions that obey the null energy condition $-1 < w < -1/3$, we see that the growth of the OTOC is considerably slower than exponential. However, for quasi-de Sitter backgrounds that might arise in models of early Universe inflation with
$w = -1 + \epsilon$ for $\epsilon \ll 1$, the growth of the OTOC can be strongly dependent on time ${\mathcal C}_k(t) \sim t^{2/\epsilon}$. Correspondingly, it seems reasonable to classify the behavior of the OTOC for these backgrounds as {\it ``quasi-chaotic''} (in the same sense that these cosmological backgrounds are often referred to as ``quasi-de Sitter'').
Crossing over to backgrounds that violate the null energy condition $w < -1$, the OTOC diverges in finite (cosmic) time, as the expanding background approaches the so-called ``big rip'' singularity.
It is not surprising to see the OTOC for these {\it ``super-chaotic''} systems display similar pathologies as the background solution as the ``big rip'' is approached.

While the discussion above in terms of the OTOC (\ref{FTOTOC2}) clearly establishes  growth at large squeezing, let's also consider the form of matrix $\hat {\cal M}$ (\ref{CosmoM}) of OTOCs and the the associated eigenvalues of the squared matrix ${\cal L}$.
At late times, the eigenvalues of $\hat {\cal M}$ become $e^{\pm\ln(a/k|\eta_0|)}\sim \left(\frac{a}{k|\eta_0|}\right)^{\pm 1}$, so that the growth of the OTOC at late times is reflected in the dominant eigenvalue of the matrix ${\cal M}$ itself. The eigenvalues of ${\cal L}$ at late times become
\be
\alpha_{\pm} \sim \left\{0,\frac{1}{2}e^{2\ln(a/k|\eta_0|)}\sin^2(\theta_0)\right\} \sim  \left\{0, \frac{1}{2} \left(\frac{a(\eta)}{k|\eta_0|}\right)^2\sin^2(\theta_0)\right\}\, .
\ee
We see the same general behavior for the dominant eigenvalue as discussed above for the OTOC (\ref{ExpandAccelOTOCLate}), confirming that analysis.

\begin{figure}[t]
\centering\includegraphics[width=.6\textwidth]{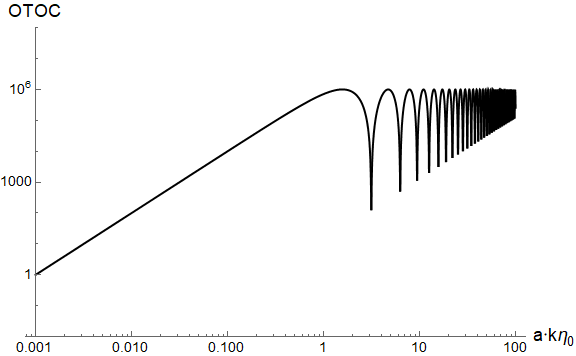}
\caption{The numerical solution to the squeezing equations of motion (\ref{squeezeEOM1})-(\ref{squeezeEOM3}) for the OTOC ${\cal C}_k$ is shown as a function of the scale factor $a$ for a radiation $w = 1/3$ background; all other decelerating backgrounds show qualitatively similar behavior. We see that the OTOC amplitude begins at a value of $1$, grows until the mode exits the horizon at $a \sim 1/(k\eta_0)^{|\beta|}$, then oscillates with a fixed amplitude, interpolating between the behaviors (\ref{DecelExpandSuperH}), (\ref{DecelExpandSubH}) described in the text.}
\label{fig:ExpandDecel_OTOC}
\end{figure}

For decelerating backgrounds $w > -1/3$, modes begin outside the horizon for sufficiently early times $a \ll 1/(k \eta_0)^{|\beta|}$; the solutions for the squeezing parameter, angle and rotation angle thus qualitatively resemble the super-horizon solutions in the accelerating case, with large and growing squeezing $r_k \gg 1$ and approximately constant angles
\be
r_k(a) & \approx & r_0+  \ln \left(a/a_0\right)\, ; \\
\phi_k(a) &\approx & -\frac{\pi}{2}\, ; \\ 
\theta_k(a) &\approx & \frac{k \eta_0 |\beta|}{(|\beta|+2)} {a^{1/|\beta|}} \ll 1\, ,
\ee
where the initial squeezing is also taken to be large $r_0 \gg 1$.
Setting $r_0 \gg 1$, $\theta_0 \approx 0$ and $\phi_0 \approx -\pi/2$, (\ref{fGeneral}) becomes
\be
f_k(\eta,\eta_0) = e^{-r_0} \left(\cosh r_k \cos(\theta_k) - \sinh r_k \cos(\theta_k+2\phi_k)\right)\, .
\label{fExpandNonAccel}
\ee
Using the solutions at some early time $\eta > \eta_0$, but still before horizon-crossing, (\ref{fExpandNonAccel}) becomes
\be
f_k(\eta,\eta_0) \approx e^{r_k-r_0} \approx \frac{a(\eta)}{a_0} \geq 1\, .
\label{DecelExpandSuperH}
\ee
We see that the OTOC ${\cal C}_k \sim f_k^2$ increases with the scale factor during these early times, up until horizon crossing.
At late times $a \gg 1/(k\eta_0)^{|\beta|}$, the modes re-enter the horizon and the squeezing ``freezes-in'', while the squeezing angle and rotation angle are large and evolving
\be
r_k(a) &\approx & r_*\, ; \\
\phi_k(a) &\approx & -\frac{3\pi}{2} + k\ \eta_0\ a^{1/|\beta|}\, ; \\
\theta_k(a) & \approx & - k \eta_0 a^{1/|\beta|}\, .
\ee
The unequal time commutator then oscillates, with an amplitude that is ``frozen-in'', set by the scale factor at horizon re-entry $a_* \sim 1/k\eta_0$
\be
f_k(\eta,\eta_0) \approx e^{-r_0} \left(\cosh r_* \cos(\theta_k) - \sinh r_* \cos(\theta_k+2\phi_k)\right) \approx \frac{a_*}{a_0} \cos\left(k\eta_0\ a^{1/|\beta|}\right) \, ,
\label{DecelExpandSubH}
\ee
where in the second approximation we assumed $r_* \gg 1$.
This behavior, as a function of the scale factor, is shown in Figure \ref{fig:ExpandDecel_OTOC}.
Combining the behavior of the unequal time commutator at early times (\ref{DecelExpandSuperH}) and late times (\ref{DecelExpandSubH}) together with the power-law behavior of the scale factor as a function of cosmic time
$a(t) = a_0 t^{|\beta|/(|\beta|+1)}$, we see that the OTOC grows as a power law ${\cal C}_k \sim t^n$ for $n<1$ at early times, then develops a strong oscillation with a saturated amplitude ${\cal C}_k \sim (a_*/a_0)^2 > 1$ set by the scale factor when the mode re-entered the horizon.

An analysis of the matrix $\hat {\cal M}$ and the eigenvalues of ${\cal L}$ tell a similar story.
At early times, for $\phi_k\approx -\pi/2$ and $\theta_k \ll 1$, the eigenvalues of ${\cal L}$ are approximately constant $\alpha_{\pm} \sim \pm 1$.
At late times, the eigenvalues oscillate with $\theta$ about fixed values
\be
\alpha_{\pm} &=& \cosh(2r_*) \sin^2(\theta_k) \pm \frac{1}{4} \sqrt{3+12\cos(2\theta_k)+ \cos(4\theta_k) + 8 \cosh(4r_*) \sin^4(\theta_k)} \sim \cosh(2r) \sin^2(\theta_k) \nonumber \\
    &\sim & e^{2r_0} \left(\frac{a_*}{a_0}\right)^2 \sin^2\left(\theta_k\right)\, ,
\ee
in rough agreement with (\ref{DecelExpandSubH}). Thus, the dominant eigenvalue
for an expanding decelerating background grows until horizon crossing, after which it saturates.

\subsection{Contracting Backgrounds}

Now we will consider contracting cosmological backgrounds with fixed equation of state.
The scale factor for a contracting universe with fixed equation of state can be written as
\begin{equation}
    a(\eta) = \left(\frac{\eta_0}{\eta}\right)^\beta = 
    \begin{cases}
    \left(\frac{\eta_0}{\eta}\right)^\beta, & 0 < \eta < \infty, ~ \eta_0 > 0 ~ \mbox{for accelerating backgrounds $\beta > 0$ ($w < -1/3$)} \\
    \left(\frac{\eta}{\eta_0}\right)^{|\beta|}, & -\infty < \eta < 0, ~ \eta_0 < 0 ~ \mbox{for decelerating backgrounds $\beta < 0$ ($w > -1/3$)}\, ,
    \end{cases}
    \label{ContractingScaleFactor}
\end{equation}
where $\beta =-2/(1+3w)$ as before.

Cosmological backgrounds that are contracting are time-reversals of their expanding counterparts, so we expect most of the qualitative features of the previous subsection to be reproduced. In particular, contracting accelerating backgrounds will be similar to expanding decelerating backgrounds, and visa versa.
The only main difference is that the scale factor for contracting backgrounds will be large at early times and become small at late times.

\begin{figure}[t]
    \centering
    \includegraphics[width=.6\textwidth]{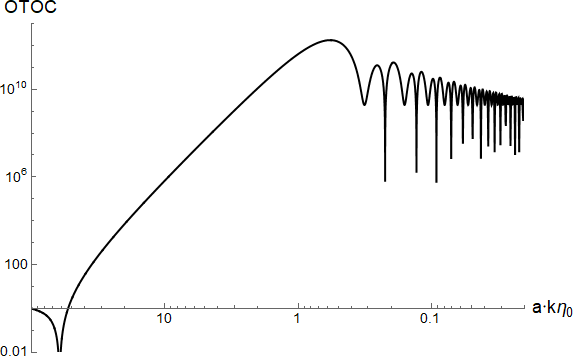}
    \caption{Numerical solutions to the squeezing equations of motion (\ref{squeezeEOM1})-(\ref{squeezeEOM3}) for the OTOC ${\cal C}_k$ show that the amplitude of the OTOC for a contracting accelerating ($w=-1$) background grows as a power of the (inverse) scale factor while the mode is outside the horizon, then saturates once the mode re-enters the horizon. Note that in contrast to the expanding accelerating case, Figure \ref{fig:dSOTOC}, the OTOC does not continue to grow exponentially at late times, even for a contracting de Sitter universe.}
    \label{fig:ContractAccelOTOC}
\end{figure}

For contracting accelerating backgrounds $w < -1/3$, modes begin outside the horizon and re-enter the horizon as the universe contracts; the corresponding solutions to the squeezing and rotation parameters are
\be
r_k(a) & \approx & r_0 -  \ln \left(a/a_0\right)\, ; \\
\phi_k(a) &\approx &  \frac{k \eta_0}{ (1+2\beta) } {a^{-1/\beta}}\, ;  \\
\theta_k(a) &\approx & -\frac{k \eta_0}{ (1+2\beta) } {a^{-1/\beta}}\, .
\ee
As with the expanding decelerating case, we have $r_0 \gg 1$, while now $\phi_0,\theta_0 \approx 0$.
Using these initial conditions in (\ref{fGeneral}), we have
\be
f_k(\eta,\eta_0) \approx e^{r_0} \left(\cosh r_k\ \cos(\theta_k) - \sinh r_k\ \cos(\theta_k+2\phi_k)\right)\, .
\label{fContractAccel}
\ee
Using the early-time squeezing solutions in (\ref{fContractAccel}), we have
\be
f_k(\eta,\eta_0) \approx e^{r_0 - r_k} \approx \frac{a}{a_0} \leq 1\, .
\ee
Note that since $a < a_0$ as the universe contracts, 
the magnitude of the OTOC ${\cal C}_k \sim f_k^2$ {\it decreases} during the early stage of a contracting accelerating universe as the scale factor decreases.
At late times, the mode re-enters the horizon around $a_* \sim 1/(k\eta_0)^\beta$, and the squeezing ``freezes-in'', while the squeezing and rotation angles become large
\be
r_k(a) &\approx & r_* = r_0 - \ln(a_*/a_0)\, ; \\
\phi_k(a) &\approx &-\pi + \frac{k\eta_0}{a^{1/\beta}}\, ; \\
\theta_k(a) &\approx & -\frac{k\eta_0}{a^{1/\beta}}\, .
\ee
The OTOC in this regime, then, oscillates with a fixed amplitude set by horizon re-entry
\be
f_k(\eta,\eta_0) \approx e^{r_0} e^{r_k}\left(\cos (\theta_k) - \cos(\theta_k + 2\phi_k)\right) \approx e^{2r_0} \frac{a_0}{a_*} \cos \left(\theta_k\right)\, .
\label{ContractAccelLate}
\ee
Altogether, for a contracting accelerating background the OTOC starts at $1$ and increases until the mode re-enters the horizon, then oscillates with a fixed amplitude $e^{2r_0} a_0/a_* \gg 1$ set by the scale factor at re-entry.
A numerical solution demonstrating this behavior is shown in Figure \ref{fig:ContractAccelOTOC}.

The eigenvalues of the matrix ${\cal L}$ tell a similar story.
At early times, the eigenvalues of ${\cal L}$ are constant $\alpha_{\pm} \sim \pm 1$, while for late times the dominant eigenvalue becomes
\be
\alpha_{\pm} &=& \cosh(2r_*) \sin^2(\theta_k) \pm \frac{1}{4} \sqrt{3+12\cos(2\theta_k)+ \cos(4\theta_k) + 8 \cosh(4r_*) \sin^4(\theta_k)} \sim \cosh(2r) \sin^2(\theta_k) \nonumber \\
    &\sim & e^{2r_0} \left(\frac{a_0}{a_*}\right)^2 \sin^2\left(\theta_k\right)\, .
\label{ContractAccelEigen}
\ee
The dominant eigenvalue is indeed approximately the square of the amplitude (\ref{ContractAccelLate}).
Altogether, the eigenvalues begin ${\cal O}(1)$, growing through horizon crossing until reaching the saturation value given in (\ref{ContractAccelEigen}). 
Note that this behavior precisely mirrors the behavior of the eigenvalues of the expanding, decelerating background from the previous subsection.
Focusing on a contracting de Sitter background, the OTOC grows proportional to the scale factor while the mode is outside the horizon, leading to a period of exponential growth of the OTOC when expressed in cosmic time. However, the OTOC soon saturates as the mode enters the horizon, leading to a \emph{transient} period of exponential growth of the OTOC, where the timescale for this period of growth is set by the horizon re-entry time.
Comparing this to the OTOC for an expanding de Sitter background, we see that the OTOC behaviors are not mirror images of each other, since the OTOC grows without saturation for the expanding case.
It would be interesting to study further whether this fundamental difference between expanding and contracting de Sitter backgrounds could be due to a difference in the validity of their respective effective field theories.

\begin{figure}[t]
    \centering
    \includegraphics[width=.6\textwidth]{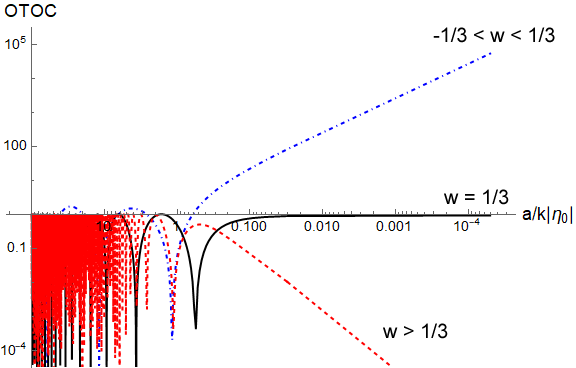}
    \caption{Numerical solutions to the squeezing equations of motion (\ref{squeezeEOM1})-(\ref{squeezeEOM3}) for the OTOC ${\cal C}_k$ show that the amplitude of the OTOC at late times (small scale factor) for a contracting decelerating universe depends on the equation of state, as discussed in (\ref{fContractDecelLate}). For equations of state $-1/3 < w < 1/3$, the OTOC grows at late times as a power of the scale factor (blue, dot-dash). For equations of state $w > 1/3$, the OTOC decays as a power of the scale factor (red dashed). For radiation backgrounds $w = 1/3$, the OTOC has a constant amplitude of one (black, solid).}
    \label{fig:ContractDecel_OTOC}
\end{figure}

Finally, let's consider a contracting decelerating cosmological background.
At early times, modes for a contracting decelerating background start deep within the horizon, with small squeezing parameter, fixed squeezing angle, and large rotation angle
\be
r_k(a) &\approx & \frac{|\beta|}{2k|\eta_0|} \frac{1}{a^{1/|\beta}}\, ; \\
\phi_k(a) & \approx & \frac{\pi}{4} + \frac{1}{4k|\eta_0|} \frac{1}{a^{1/|\beta|}}\, ; \\
\theta_k&\approx & -k|\eta_0|\ a^{1/|\beta|}\, .
\ee
Using $r_0 \ll 1$ and $\phi_0 \approx \pi/4$, the magnitude of the OTOC takes the general form
\be
f_k(\eta,\eta_0) \approx \cosh r_k\ \cos(\theta_k-\theta_0) - \sinh r_k\ \cos(\theta_k-\theta_0+2\phi_k)\, .
\label{fContractDecel}
\ee
The corresponding magnitude of the OTOC at early times thus oscillates with an ${\cal O}(1)$ amplitude
\be
f_k(\eta,\eta_0) \approx \cos\left(\theta_k - \theta_0\right) \sim {\cal O}(1)\, .
\label{fContractDecelEarly}
\ee
At late times (again, corresponding to small scale factor), the modes exit the horizon, and the squeezing parameter begins to grow, while the squeezing and rotation angles ``freeze-out'' to zero.
Unlike the previous cases, however, we need to keep careful track of how the squeezing angle $\phi_k$ decays as a function of the scale factor and the equation of state (controlled by $\beta = -2/(1+3w)$)
\be
r_k &\approx & - \ln (a/a_0)\, ; \label{ContractDecelLateR}\\
\phi_k &\approx & \begin{cases}
    \frac{k \eta_0}{1-2|\beta|} a^{1/|\beta|} & \mbox{ for } \beta < -1/2,\ (-1/3 < w < 1) \cr
    B a^{1/|\beta|} - \frac{k\eta_0}{|\beta|} a^{1/|\beta|} a^{1/|\beta|} \ \ln a & \mbox{ for } \beta = -1/2,\ (w=1) \cr
    B a^2 & \mbox{ for } -1/2 < \beta < 0,\ (w > 1)
\end{cases}\, ; \label{ContractDecelLatePhi}\\
\theta_k &\approx & 0 + {\cal O}\left(a^{1/|\beta|}\right)\, . \label{ContractDecelLateTheta}
\ee
As discussed in \cite{us2} the behavior of the solution for the scale factor changes with $\beta$ depending on whether the homogeneous or inhomogeneous term dominates the right-hand side of the $\phi_k$ equation of motion. The constant $B$ is determined by the initial conditions for $\phi_k$; its precise value will not be important for us to determine the qualitative behavior of the OTOC from (\ref{ContractDecelLatePhi}).
Inserting the solutions (\ref{ContractDecelLateR})-(\ref{ContractDecelLateTheta}) into (\ref{fContractDecel}), we have (before inserting specific behaviors of $\phi_k$)
\be
f_k(a,a_0) &\approx & e^{-r_k} \cos(\theta_0) + e^{r_k} \sin(\theta_0) \sin(2\phi_k) \\
&\approx & \left(\frac{a}{a_0}\right) \cos(\theta_0) + 2\left(\frac{a_0}{a}\right) \phi_k(a)\ \sin(\theta_0)\, .
\label{fContractLateApprox}
\ee
Notice that the first term of (\ref{fContractLateApprox}) is always decreasing as $a \rightarrow 0$ at late times for a contracting background. However, depending on the functional form of $\phi_k(a)$, the second term of (\ref{fContractLateApprox}) could be either increasing or decreasing with time.
In particular, as long as the squeezing angle scales with a power of the scale factor less than one $\phi_k(a) \sim a^n$ for $n < 1$, then the second term of (\ref{fContractLateApprox}) will \emph{increase} with time and dominate at late times. From (\ref{ContractDecelLatePhi}), the power of the squeezing angle scales with a power of the scale factor less than one for $|\beta| > 1$, which translates to an equation of state of $-1/3 < w < 1/3$. For these equations of state, which encompass most known large-scale matter fluids, the amplitude of the OTOC at late times will grow with time. On the other hand, for $|\beta| < 1$, corresponding to $w > 1/3$, the amplitude of the OTOC will decay with time; the marginal case $|\beta| = 1$ (w = 1/3) will result in a constant amplitude at late times
\be
f_k(a,a_0) \approx \begin{cases}
    \left(\frac{a}{a_0}\right) \cos(\theta_0) & \mbox{ for } |\beta| < 1\ (w > 1/3) \cr
    2 k \eta_0 a_0 \sin(\theta_0) & \mbox{ for } |\beta| = 1\ (w = 1/3) \cr
    2 \frac{k \eta_0}{1-2|\beta|} \sin(\theta_0) \frac{a_0}{a^{1-1/|\beta}} & \mbox{ for } |\beta| > 1 (-1/3 < w < 1/3)\, .
\end{cases}
\label{fContractDecelLate}
\ee
We can clearly see this transition in the late-time behavior as a function of the equation of state in the numerical solutions displayed in Figure \ref{fig:ContractDecel_OTOC}.

As with the other contracting case, we should re-examine this result for the OTOC by calculating the eigenvalues of the matrix ${\cal L}$.
At early times, the eigenvalues of the matrix ${\cal L}$ are ${\cal O}(1)$, and given by $\alpha_{\pm} = \{-\cos(2(\theta_k-\theta_0)),1\}$.
At late times, the eigenvalues become
\be
\alpha_{\pm} = \cosh r_k \sin(\theta_k-\theta_0) \pm \sqrt{-1+\cosh^2(r_k) \sin^2(\theta_k-\theta_0)} \approx \left\{0,\frac{1}{2} e^{2r_k}\sin^2(\theta_0)\right\} \approx \left\{0,2\left(\frac{a_0}{a}\right)^2 \sin^2(\theta_0)\right\}\, .
\ee
From these eigenvalues we see dramatically different behavior from that of the single OTOC (\ref{fContractDecelLate}): while the late-time behavior of the OTOC (\ref{fContractDecelLate}) depended on the equation of state, the behavior of the eigenvalues shows a dominant growing mode as the universe contracts.
This illustrates a shortcoming of focusing only on a single OTOC.
By considering the spectrum of eigenvalues from the OTOC matrix ${\cal L}$, we see a more complete picture of the behavior of the OTOC that is independent of the particular choice of field variables inserted into the OTOC.

Remarkably, we see that the eigenvalues of ${\cal L}$ for the contracting decelerating backgrounds match those for the expanding accelerating backgrounds at late times!
In fact, if we chose to calculate the OTOC based on $[\hat v_k(\eta_0),\hat \pi_k(\eta)]$, with the roles of $\eta\leftrightarrow \eta_0$ switched, for a contracting decelerating background, then we should recover the single OTOC results from the expanding accelerating background, illustrating the incompleteness of studying an OTOC based on only one particular combination of fields.
Note that even while the eigenvalues of ${\cal L}$ for the OTOC are growing at late times, the scale factor, written in terms of cosmic time $t$, does not exhibit exponential behavior.
In particular, for the contracting decelerating backgrounds considered here, the scale factor written in cosmic time are power-law functions of $t$, and so are \emph{sub-chaotic}.

Overall, we have seen a general universal behavior in the OTOC (and its generalization to the dominant eigenvalue of ${\cal L}$) for all cosmological backgrounds: When the mode is inside the horizon, the OTOC has a fixed amplitude, while modes outside the horizon grow proportional to the scale factor (for expanding backgrounds) or its inverse (for contracting backgrounds).

\section{Discussion}
\label{sec:Discussion}

Tools and techniques from quantum information theory, such as the \emph{out-of-time-order correlator} (OTOC) or circuit complexity, can provide alternative perspectives of interesting quantum systems.
The OTOC can be a useful probe of the time-dependent overlap between two operators, and has a natural interpretation in the semiclassical limit as the separation between two nearby trajectories in phase space.
Analogous to the classical case, an exponentially growing OTOC is commonly associated with quantum chaos (although not all systems with quantum chaos exhibit an exponential dependence).
In this paper, we studied the OTOC for squeezed states and their application in the time-evolution of cosmological perturbations.

We find that the OTOC for generic squeezed states can be calculated in closed form and is independent of the averaging procedure.
Highly squeezed states, characterized by the squeezing parameter $r$, have an exponentially large OTOC, ${\cal C}\sim e^{2r}$, so that any linear dependence of the squeezing on time leads to exponential growth. In this sense, a highly squeezed state is ``primed'' for quantum chaos.
However, we found that an over-reliance on a single combination of fields in the OTOC (such as the canonical choice $[\hat q(\eta),\hat p(\eta_0)]$) can mask general behavior of the system. Instead, one should construct the squared matrix ${\cal L}_{ij}$ constructed from the symplectic matrix $\hat{\cal M}_{ij}$ of unequal-time commutators, analogously to the classical case.
The eigenvalues of ${\cal L}_{ij}$ capture the general OTOC behavior of the system, and their late-time exponential behavior can be used as a more robust diagnostic of quantum chaos.
We applied our formalism of squeezed states to calculate the OTOC of the inverted harmonic oscillator, finding an exponential growth of both the canonical OTOC and the eigenvalues of ${\cal L}$ at late times.
It is interesting to compare this result to other diagnostic probes of the inverted harmonic oscillator. For example, the single-evolved complexity \cite{us} does not show any growth at late times consistent with quantum chaos, while the double-evolved complexity \cite{me1} and displacement operator complexity \cite{Bhattacharyya:2020art} does show such growth.
Thus the OTOC, particularly the eigenvalues of the squared OTOC matrix ${\cal L}_{ij}$, may be among the more sensitive measures of time-dependent dynamics.

After generalizing our techniques for computing the OTOC to two-mode squeezed states of continuous Fourier modes, we studied the OTOC for cosmological perturbations on expanding and contracting backgrounds of fixed equation of state.
Previous studies of the complexity of cosmological perturbations \cite{us,us2} uncovered a rich structure in its time-dependent behavior, including surprising bounds on the growth rate of complexity reminiscent of \cite{Maldacena:2015waa}.
Here, we find that only \emph{expanding de Sitter space} leads to an exponential growth of the OTOC at late (cosmic) times, suggesting that only expanding de Sitter can be described as quantum chaotic at late times. 
In Appendix \ref{sec:dSOTOC} we further used the exact mode function solution for de Sitter to calculate the OTOC in both Fourier space and position space exactly, the former showing excellent agreement with the squeezed state result.
In position space, the OTOC between two operators is only non-zero inside their causal lightcone, as expected, and it grows exponentially in a similar way as the Fourier space OTOC.
Altogether, the OTOC for cosmological perturbations on an expanding de Sitter background 
shows exponential growth
as other OTOC calculations done using different techniques \cite{Shiu}, and leads to an identical putative Lyapunov exponent.

While expanding de Sitter is of particular interest, because of its applications to early- and late-time acceleration of our Universe as well as its theoretical similarity to black hole spacetimes, the squeezed state language makes it easy to include in our analysis expanding and contracting backgrounds with arbitrary fixed equation of state.
The OTOC for expanding accelerating backgrounds begins small while modes are still within the horizon, then grows after horizon exit. As noted before, only for de Sitter backgrounds is the resulting growth of the OTOC exponential. For expanding accelerating backgrounds that obey the null energy condition $-1 < w < -1/3$, the OTOC grows as a ``sub-chaotic'' power of cosmic time, while the OTOC for backgrounds that violate the null energy condition $w < -1$ has a ``super-chaotic'' growth, diverging in finite time.
The OTOC for expanding decelerating backgrounds shows a qualitatively different behavior, however, as it grows initially, then ``freezes in'' once the mode enters the horizon.
For expanding backgrounds the dominant eigenvalue of the squared matrix ${\cal L}_{ij}$ matches qualitatively the behavior of the canonical OTOC.

We expect that the behavior of the OTOC for contracting backgrounds should mirror their expanding counterparts, namely that a contracting accelerating background should behave similar to an expanding decelerating background, and vice versa.
Indeed, we find for contracting accelerating backgrounds, including a contracting de Sitter universe, the OTOC grows initially before ``freezing-in'' once the mode enters the horizon.
Since the growth of the OTOC is proportional to the scale factor, this means that the OTOC for a contracting de Sitter universe has a transient period of exponential growth, before saturating at a fixed value.
However, the behavior of the canonical OTOC for contracting decelerating backgrounds does not directly mirror that of an expanding accelerating background; instead of uniform growth after exiting the horizon, the late-time behavior of the OTOC qualitatively changes as a function of the equation of state. For equations of state less than that of radiation $-1/3 < w < 1/3$, including pressureless matter, the OTOC grows at late time as a power of the scale factor. For an equation of state equal to that of radiation $w = 1/3$ the OTOC is constant and equal to $1$ at late times, while for equations of state ``stiffer'' than radiation $w > 1/3$ the OTOC decays at late times.
Interestingly, while the behavior of the canonical OTOC for a contracting decelerating background does not mirror that of its expanding accelerating counterpart, the eigenvalues of the squared matrix ${\cal L}_{ij}$ have precisely the same qualitative form for both backgrounds.
This is further evidence that reliance on a single OTOC as a probe of the dynamics of a quantum system can mislead, and that  important general features can be better extracted from the squared matrix inspired by classical chaos.
Note that while a contracting decelerating background has a growing OTOC at late times, the growth is always slower than exponential, so that it is ``sub-chaotic'' in this sense.
Overall, we find a universal behavior for the generalization of the OTOC to the eigenvalues of the squared matrix ${\cal L}$: the OTOC oscillates with a fixed amplitude for modes inside the horizon, while it grows as the scale factor (or its inverse) for modes outside the horizon, for expanding and contracting backgrounds respectively.
Note that this means that only de Sitter backgrounds, either expanding or contracting, have a period in which the OTOC grows exponentially with cosmic time.

It is interesting to compare our results here for the OTOC of cosmological backgrounds to the corresponding results for quantum circuit complexity found in \cite{us,us2}.
Both the OTOC and the circuit complexity are expected to be probes of time-dependent dynamics of a quantum system, and can potentially signal the onset of quantumm chaos.
In \cite{us,us2}, the complexity grows at late times for expanding accelerating (and contracting decelerating) backgrounds.
While we did find that the OTOC for contracting decelerating backgrounds with $w > 1/3$ decreases at late times, the eigenvalues of the squared matrix ${\cal L}_{ij}$ increase at late times in an identical way to expanding accelerating backgrounds.
Remarkably the slope of the growth of complexity was found to saturate for equations of state less than $-5/3$ for expanding backgrounds (or greater than $1$ for contracting backgrounds), evoking echos of the bound on the Lyapunov exponent \cite{Maldacena:2015waa}.
We do not see any corresponding saturation in the growth or growth rate of the OTOC for expanding accelerating (or contracting decelerating) backgrounds, suggesting that circuit complexity is probing slightly different features of the quantum system.
Further, for expanding decelerating (and contracting accelerating) backgrounds, \cite{us,us2} found that the circuit complexity actually decreases with time until ``freezing in'' after entering the horizon, while we find here the opposite behavior that the OTOC increases until entering the horizon.
It would be interesting to study whether there is a deeper reason these two diagnostics give qualitatively different behaviors, and whether there is an analogue of the squared eigenvalues of the OTOC for circuit complexity.

While we have seen that only an expanding de Sitter background leads to exponential growth at late times, suggesting that expanding de Sitter space experiences quantum chaos, it is less clear how to see a separation of dissipation and scrambling time scales in this system.
Thus, it is not immediately clear what an interpretation of the OTOC for a microscopic theory of the de Sitter horizon would be.
Indeed, for eternal de Sitter the growth of the OTOC continues without bound, which is inconsistent with broad expectations of the OTOC at very late times.
Perhaps general expectations that the OTOC saturate at very late times imply the breakdown of the
effective field theory used here, and could lead to a cap on the amount of de Sitter expansion, similar to other proposals \cite{Brahma:2020zpk} (although the curvature perturbation is fixed on superhorizon scales, reducing the likelihood of gravitational backreaction).
Additionally, since de Sitter is the fastest ``scrambler'' among backgrounds that satisfy the null energy condition, with the fastest growth of the OTOC, this may make it a natural cosmological background for our Universe, since we have only one Universe (that we know of) and it appears to have early-time and late-time (quasi)-de Sitter expansion periods. We leave these and related questions for future work.

\section*{Acknowledgements}
We would like to thank Arpan Bhattacharyya, Pawel Caputa, Sayura Das, Jeff Murugan, Dario Rosa and Bin Yan for helpful conversations and discussions. S.H. would like to thank the University of Cape Town for funding this project. 

\appendix

\section{OTOC for de Sitter}
\label{sec:dSOTOC}

In Sections \ref{sec:Squeezed} and \ref{sec:CosmoOTOC}, we investigated the OTOC for Fourier modes of cosmological perturbations using the squeezed state formalism, which allowed us to easily investigate several different cosmological solutions using a uniform formalism. However, there are some cosmological backgrounds, such as de Sitter, where the Fourier mode functions are known precisely in closed form. It can be advantageous to perform explicit calculations using these closed form expressions for this specific and important background.

Following the analysis of Section \ref{sec:Squeezed}, the Hamiltonian for Fourier modes in de Sitter space can be written as in (\ref{CosmoH}), with $z'/z = a'/a = -1/\eta$. The mode functions $\hat v_{\vec k}(\eta)$ and their corresponding momenta $\hat \pi_{\vec k}(\eta)$ for a mode that begins in the ground state in the far past $\eta_i \rightarrow -\infty$ can be written
\be
\hat v_{\vec k}(\eta) &=& v_{\vec{k}}(\eta)\ \hat a_{\vec{k}}(\eta_i) + v^*_{-\vec{k}}\ \hat a^\dagger_{-\vec{k}}(\eta_i)\, ; \\
\hat \pi_{\vec{k}}(\eta) &=& \dot v_{-\vec{k}}(\eta)\ \hat a_{-\vec{k}}(\eta_i) + \dot v_{\vec{k}}^*(\eta)\ \hat a_{\vec{k}}^\dagger(\eta_i)\, ,
\ee
where
\be
v_{\vec{k}}(\eta) = \frac{1}{\sqrt{2k}} \left(1-\frac{i}{k\eta}\right) e^{-ik\eta}\, .
\ee
Notice that in the far past $\eta_i \rightarrow -\infty$ this becomes the flat space ground state $v_{\vec{k}} \sim e^{-ik\eta}/\sqrt{2k}$, as expected.
For normalized raising and lowering operators, the equal-time commutator between the mode $\hat v_{\vec{k}}$ and its momentum $\hat \pi_{\vec{k}}$ gives the canonical result
\be
\left[\hat v_{\vec{k}}(\eta),\hat\pi_{\vec{k}'}(\eta)\right] = i (2\pi)^3 \delta^3\left(\vec{k}-\vec{k}'\right)\, .
\ee
The unequal time commutator between these operators, evaluated at $\eta$ and $\eta'$, gives a more interesting result
\be
\left[\hat v_{\vec{k}}(\eta),\hat\pi_{\vec{k}'}(\eta')\right] = i (2\pi)^3 \delta^3\left(\vec{k}-\vec{k}'\right) f_k(\eta,\eta')\, ,
\ee
where the amplitude function $f_k(\eta,\eta')$ is
\be
f_k(\eta,\eta') = \cos(k(\eta-\eta')) - \frac{\sin(k(\eta-\eta'))}{k} \left(\frac{1}{\eta} - \frac{1}{\eta'}\right) + \frac{\cos(k(\eta-\eta'))}{k^2} \frac{1}{\eta'} \left(\frac{1}{\eta} - \frac{1}{\eta'}\right) + \frac{\sin(k(\eta-\eta'))}{k^3} \frac{1}{\eta\, \eta^{'2}}\, .
\label{fdS}
\ee
Notice that for $\eta\rightarrow \eta'$, $f_k\rightarrow 1$, recovering the equal-time result.

The OTOC, written as the thermally averaged square of the unequal time commutator
\be
{\mathcal C}_{\vec{k}}(\eta) \equiv - \langle [\hat v_{\vec{k}_1}(\eta),\hat \pi_{\vec{k}_2}(\eta_0)][\hat v_{\vec{k}_3}(\eta),\hat \pi_{\vec{k}_4}(\eta_0)]\rangle_{\beta}\, ,
\ee
again simplifies to the square of the commutator
\be
{\mathcal C}_{\vec{k}}(\eta) = - [\hat v_{\vec{k}_1}(\eta),\hat \pi_{\vec{k}_2}(\eta_0)][\hat v_{\vec{k}_3}(\eta),\hat \pi_{\vec{k}_4}(\eta_0)]\, ,
\ee
because the commutator is a $c$-number, 
so that the amplitude of the OTOC in Fourier space is given by the square of the amplitude
${\cal C}_{\vec{k}}(\eta) \sim \left(f_k(\eta,\eta')\right)^2$.
Taking our initial time to be in the far past $\eta' \rightarrow -\infty$, (\ref{fdS}) simplifies to become
\be
f_k(\eta,\eta') \approx \cos(k(\eta-\eta')) - \frac{\sin(k(\eta-\eta'))}{k\eta}\,.
\ee
At late times $\eta\rightarrow 0^-$, this is dominated by the second term $f_k(\eta,\eta') \approx \sin(k\eta')/(k\eta)$, so that the amplitude of the OTOC is given by
\be
{\cal C}_k(\eta) \sim \frac{\sin^2(k\eta')}{(k\eta)^2}\sim \left(\frac{H_{dS}}{k}\right)^2 \sin^2(k\eta')\ e^{2H_{dS}t}\, ,
\label{dSOTOCFourier}
\ee
where we converted from conformal time $\eta$ to cosmic time $t$. Up to ${\cal O}(1)$ factors, we find the same results for de Sitter space as (\ref{ExpandAccelOTOCLate}) in Section \ref{sec:CosmoOTOC} using the squeezed state formalism.
In particular, we find that the OTOC for de Sitter space exhibits exponential growth at late times ${\cal C}_k \sim e^{2\lambda t}$, with a putative Lyapunov exponent given by the de Sitter Hubble constant $\lambda = H_{dS}$.

\begin{figure}[t]
    \centering
    \includegraphics[width=.5\textwidth]{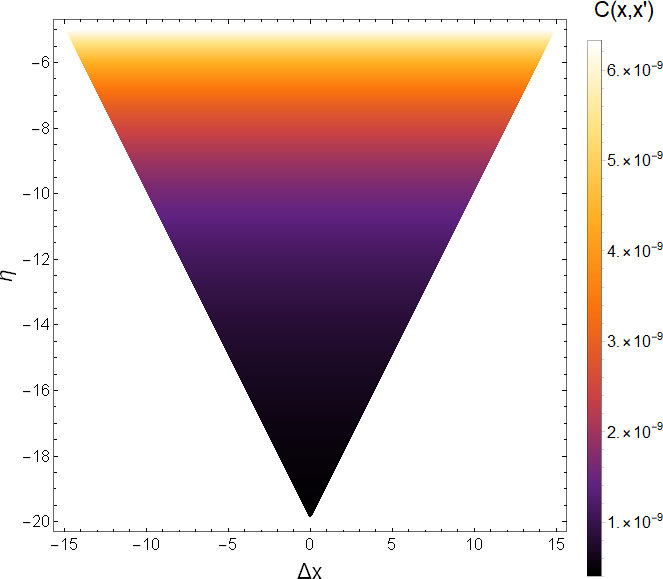}
    \caption{The position-space OTOC ${\cal C}(x,x')$ for cosmological perturbations in de Sitter space (\ref{dSOTOCPosition}) between the field operator $\hat v(\vec{x},\eta)$ and its conjugate momentum $\hat \pi(\vec{x}',\eta')$, shown here in conformal coordinates $\eta,x$, grows as $1/\eta^2$ inside the lightcone (we have taken $\eta' = -20$ for concreteness).}
    \label{fig:my_label}
\end{figure}

While we have been focused on the Fourier space behavior of the OTOC, it is also potentially interesting to consider the position space unequal time commutator
\be
\left[\hat v(\vec{x},\eta),\hat \pi(\vec{x}',\eta')\right] = \int \frac{d^3k\, d^3k'}{(2\pi)^6} \left[\hat v_{\vec{k}}(\eta),\hat \pi_{\vec{k}'}(\eta')\right] e^{i\vec{k}\cdot \vec{x} - i \vec{k}'\cdot \vec{x}'} = i\int \frac{d^3k}{(2\pi)^3}\ f_k(\eta,\eta')\ e^{i\vec{k}\cdot(\vec{x}-\vec{x}')}\, .
\ee
Using the expression (\ref{fdS}) for the amplitude of the Fourier OTOC, the result becomes
\be
\left[\hat v(\vec{x},\eta),\hat \pi(\vec{x}',\eta')\right] = 
    \frac{i}{2\pi^2} \left[-2\pi \Delta \eta \frac{\delta(\Delta \eta^2-\Delta x^2)}{\Delta \eta^2-\Delta x^2} + \pi \left(\frac{1}{\eta}-\frac{1}{\eta'}\right) \delta(\Delta\eta^2-\Delta x^2) - \frac{\pi}{2} \frac{1}{\eta\eta^{'2}} \mbox{sign}(\Delta \eta) \Theta(\Delta \eta-\Delta x)\right]\, ,
\ee
where $\Delta \eta = \eta'-\eta$, $\Delta x = x'-x$ and the Heaviside function $\Theta(x)$ in the last term only gives a non-zero contribution inside the lightcone.
As expected, the unequal-time commutator vanishes outside the lightcone, reflecting causality. 
Taking the position-space OTOC as the square of the thermally averaged unequal time commutator, the thermal averaging is trivial
\be
{\cal C}(x,x') \equiv -i \langle \left[\hat v(\vec{x},\eta),\hat \pi(\vec{x}',\eta')\right]^2 \rangle_{\beta} = -i \left[\hat v(\vec{x},\eta),\hat \pi(\vec{x}',\eta')\right]^2\, ,
\ee
and so working strictly within the lightcone, we see the familiar exponential growth of the OTOC in de Sitter space
\be
{\cal C}(x,x') = \frac{1}{(2\pi)^4} \frac{1}{\eta'^{4}} \frac{1}{\eta^2} = \frac{1}{(4\pi)^2\eta'^4} H_{dS}^2\ e^{2H_{dS}t}\, ,
\label{dSOTOCPosition}
\ee
where we again used the relationship between conformal time and cosmic time.
The position-space OTOC (\ref{dSOTOCPosition}) reflects a similar structure as we found in Fourier space (\ref{dSOTOCFourier}), including exponential growth with the de Sitter Hubble constant, and a factor of $H_{dS}^2$.
Interestingly, if we take our initial time to be past infinity $\eta'\rightarrow -\infty$ the position-space OTOC (\ref{dSOTOCPosition}) vanishes for points inside the lightcone, and is non-zero only on the lightcone itself, reflecting a lack of correlations between the momentum at past infinity and the field at any finite time not on the lightcone.
Similar results hold for the field-field and momentum-momentum unequal-time commutators and their associated position-space OTOCs.
It would be interesting to explore how the behavior of the de Sitter OTOC near the lightcone relate to the bounds presented in \cite{Mezei_2020}.

\bibliographystyle{utphysmodb}

\bibliography{refs}

\providecommand{\href}[2]{#2}\begingroup\raggedright\begin{thebibliography}{10}

\bibitem{Kitaev2015}
A.~Kitaev,  {\em A simple model of quantum holography}, Proceedings of the KITP
  Program: Entanglement in Strongly-Correlated Quantum Matter, (Kavli Institute
  for Theoretical Physics, Santa Barbara) {\bf Vol. 7} (2015).

\bibitem{Larkin1969}
A.~I. {Larkin} and Y.~N. {Ovchinnikov},  {\em {Quasiclassical Method in the
  Theory of Superconductivity}}, Soviet Journal of Experimental and Theoretical
  Physics {\bf 28} (June, 1969) 1200.

\bibitem{Maldacena:2015waa}
J.~Maldacena, S.~H. Shenker and D.~Stanford,  {\em {A bound on chaos}}, JHEP
  {\bf 08} (2016) 106 [\href{http://www.arXiv.org/abs/1503.01409}{{\tt
  1503.01409}}].

\bibitem{Hashimoto:2017oit}
K.~Hashimoto, K.~Murata and R.~Yoshii,  {\em {Out-of-time-order correlators in
  quantum mechanics}}, JHEP {\bf 10} (2017) 138
  [\href{http://www.arXiv.org/abs/1703.09435}{{\tt 1703.09435}}].

\bibitem{Rozenbaum_2017}
E.~B. Rozenbaum, S.~Ganeshan and V.~Galitski,  {\em Lyapunov Exponent and
  Out-of-Time-Ordered Correlator’s Growth Rate in a Chaotic System}, Physical
  Review Letters {\bf 118} (Feb, 2017).

\bibitem{Rozenbaum_2019}
E.~B. Rozenbaum, S.~Ganeshan and V.~Galitski,  {\em Universal level statistics
  of the out-of-time-ordered operator}, Physical Review B {\bf 100} (Jul,
  2019).

\bibitem{Gharibyan_2019}
H.~Gharibyan, M.~Hanada, B.~Swingle and M.~Tezuka,  {\em Quantum Lyapunov
  spectrum}, Journal of High Energy Physics {\bf 2019} (Apr, 2019).

\bibitem{bhattacharyya2019web}
A.~Bhattacharyya, W.~Chemissany, S.~S. Haque and B.~Yan,  {\em Towards the Web
  of Quantum Chaos Diagnostics}, 2019.

\bibitem{Kudler-Flam:2019kxq}
J.~Kudler-Flam, L.~Nie and S.~Ryu,  {\em {Conformal field theory and the web of
  quantum chaos diagnostics}}, JHEP {\bf 01} (2020) 175
  [\href{http://www.arXiv.org/abs/1910.14575}{{\tt 1910.14575}}].

\bibitem{QuantLyapunovSpectrum}
H.~Gharibyan, M.~Hanada, B.~Swingle and M.~Tezuka,  {\em {Quantum Lyapunov
  Spectrum}}, JHEP {\bf 04} (2019) 082
  [\href{http://www.arXiv.org/abs/1809.01671}{{\tt 1809.01671}}].

\bibitem{me3}
T.~Ali, A.~Bhattacharyya, S.~S. Haque, E.~H. Kim, N.~Moynihan and J.~Murugan,
  {\em {Chaos and Complexity in Quantum Mechanics}}, Phys. Rev. {\bf D101}
  (2020), no.~2, 026021
[\href{http://www.arXiv.org/abs/1905.13534}{{\tt 1905.13534}}].

\bibitem{Bhattacharyya:2020art}
A.~Bhattacharyya, W.~Chemissany, S.~S. Haque, J.~Murugan and B.~Yan,  {\em {The
  Multi-faceted Inverted Harmonic Oscillator: Chaos and Complexity}},
  \href{http://www.arXiv.org/abs/2007.01232}{{\tt 2007.01232}}.

\bibitem{Blume_Kohout_2003}
R.~Blume-Kohout and W.~H. Zurek,  {\em Decoherence from a chaotic environment:
  An upside-down “oscillator” as a model}, Physical Review A {\bf 68} (Sep,
  2003).

\bibitem{Morita_2019}
T.~Morita,  {\em Thermal Emission from Semiclassical Dynamical Systems},
  Physical Review Letters {\bf 122} (Mar, 2019).

\bibitem{bueno2019complexity}
P.~Bueno, J.~M. Magan and C.~S. Shahbazi,  {\em Complexity measures in QFT and
  constrained geometric actions}, 2019.

\bibitem{Yan_2020}
B.~Yan, L.~Cincio and W.~H. Zurek,  {\em Information Scrambling and Loschmidt
  Echo}, Physical Review Letters {\bf 124} (Apr, 2020).

\bibitem{Betzios_2016}
P.~Betzios, N.~Gaddam and O.~Papadoulaki,  {\em The black hole S-Matrix from
  quantum mechanics}, Journal of High Energy Physics {\bf 2016} (Nov, 2016).

\bibitem{Hegde_2019}
S.~S. Hegde, V.~Subramanyan, B.~Bradlyn and S.~Vishveshwara,  {\em Quasinormal
  Modes and the Hawking-Unruh Effect in Quantum Hall Systems: Lessons from
  Black Hole Phenomena}, Physical Review Letters {\bf 123} (Oct, 2019).

\bibitem{Hashimoto_2020}
K.~Hashimoto, K.-B. Huh, K.-Y. Kim and R.~Watanabe,  {\em Exponential growth of
  out-of-time-order correlator without chaos: inverted harmonic oscillator},
  Journal of High Energy Physics {\bf 2020} (Nov, 2020).

\bibitem{Hayden_2007}
P.~Hayden and J.~Preskill,  {\em Black holes as mirrors: quantum information in
  random subsystems}, Journal of High Energy Physics {\bf 2007} (Sep, 2007)
  120–120.

\bibitem{Sekino_2008}
Y.~Sekino and L.~Susskind,  {\em Fast scramblers}, Journal of High Energy
  Physics {\bf 2008} (Oct, 2008) 065–065.

\bibitem{us}
A.~Bhattacharyya, S.~Das, S.~Shajidul~Haque and B.~Underwood,  {\em
  {Cosmological Complexity}}, Phys. Rev. D {\bf 101} (2020), no.~10, 106020
  [\href{http://www.arXiv.org/abs/2001.08664}{{\tt 2001.08664}}].

\bibitem{us2}
A.~Bhattacharyya, S.~Das, S.~S. Haque and B.~Underwood,  {\em {The Rise of
  Cosmological Complexity: Saturation of Growth and Chaos}}, Phys. Rev. Res.
  {\bf 2} (2020), no.~3, 033273
  [\href{http://www.arXiv.org/abs/2005.10854}{{\tt 2005.10854}}].

\bibitem{ChoudhuryOTOC}
S.~Choudhury,  {\em {The Cosmological OTOC: Formulating new cosmological
  micro-canonical correlation functions for random chaotic fluctuations in
  Out-of-Equilibrium Quantum Statistical Field Theory}}, Symmetry {\bf 12}
  (2020) 1527 [\href{http://www.arXiv.org/abs/2005.11750}{{\tt 2005.11750}}].

\bibitem{ChoudhuryComplexity}
P.~Bhargava, S.~Choudhury, S.~Chowdhury, A.~Mishara, S.~P. Selvam, S.~Panda and
  G.~D. Pasquino,  {\em {Quantum aspects of chaos and complexity from bouncing
  cosmology: A study with two-mode single field squeezed state formalism}},
  \href{http://www.arXiv.org/abs/2009.03893}{{\tt 2009.03893}}.

\bibitem{Grishchuk}
L.~P. Grishchuk and Y.~V. Sidorov,  {\em Squeezed quantum states of relic
  gravitons and primordial density fluctuations}, Phys. Rev. D {\bf 42} (Nov,
  1990) 3413--3421.

\bibitem{Albrecht}
A.~Albrecht, P.~Ferreira, M.~Joyce and T.~Prokopec,  {\em {Inflation and
  squeezed quantum states}}, Phys. Rev. {\bf D50} (1994) 4807--4820
[\href{http://www.arXiv.org/abs/astro-ph/9303001}{{\tt astro-ph/9303001}}].

\bibitem{Martin1}
J.~Martin,  {\em {Inflationary perturbations: The Cosmological Schwinger
  effect}}, Lect. Notes Phys. {\bf 738} (2008) 193--241
[\href{http://www.arXiv.org/abs/0704.3540}{{\tt 0704.3540}}].

\bibitem{Martin2}
J.~Martin,  {\em {Cosmic Inflation, Quantum Information and the Pioneering Role
  of John S Bell in Cosmology}}, Universe {\bf 5} (2019), no.~4, 92
[\href{http://www.arXiv.org/abs/1904.00083}{{\tt 1904.00083}}].

\bibitem{BellCMB1}
J.~Martin and V.~Vennin,  {\em {Quantum Discord of Cosmic Inflation: Can we
  Show that CMB Anisotropies are of Quantum-Mechanical Origin?}}, Phys. Rev. D
  {\bf 93} (2016), no.~2, 023505
  [\href{http://www.arXiv.org/abs/1510.04038}{{\tt 1510.04038}}].

\bibitem{BellCMB2}
J.~Martin and V.~Vennin,  {\em {Obstructions to Bell CMB Experiments}}, Phys.
  Rev. D {\bf 96} (2017), no.~6, 063501
  [\href{http://www.arXiv.org/abs/1706.05001}{{\tt 1706.05001}}].

\bibitem{Shiu}
L.~Aalsma and G.~Shiu,  {\em {Chaos and complementarity in de Sitter space}},
  JHEP {\bf 05} (2020) 152 [\href{http://www.arXiv.org/abs/2002.01326}{{\tt
  2002.01326}}].

\bibitem{Ando}
K.~Ando and V.~Vennin,  {\em {Bipartite temporal Bell inequalities for two-mode
  squeezed states}}, \href{http://www.arXiv.org/abs/2007.00458}{{\tt
  2007.00458}}.

\bibitem{Mukhanov}
V.~F. Mukhanov, H.~Feldman and R.~H. Brandenberger,  {\em {Theory of
  cosmological perturbations. Part 1. Classical perturbations. Part 2. Quantum
  theory of perturbations. Part 3. Extensions}}, Phys. Rept. {\bf 215} (1992)
  203--333.

\bibitem{Gibbons:1977mu}
G.~Gibbons and S.~Hawking,  {\em {Cosmological Event Horizons, Thermodynamics,
  and Particle Creation}}, Phys. Rev. D {\bf 15} (1977) 2738--2751.

\bibitem{Geng:2020kxh}
H.~Geng,  {\em {Non-local Entanglement and Fast Scrambling in De-Sitter
  Holography}}, \href{http://www.arXiv.org/abs/2005.00021}{{\tt 2005.00021}}.

\bibitem{me1}
T.~Ali, A.~Bhattacharyya, S.~Shajidul~Haque, E.~H. Kim and N.~Moynihan,  {\em
  {Time Evolution of Complexity: A Critique of Three Methods}}, JHEP {\bf 04}
  (2019) 087
[\href{http://www.arXiv.org/abs/1810.02734}{{\tt 1810.02734}}].

\bibitem{Brahma:2020zpk}
S.~Brahma, O.~Alaryani and R.~Brandenberger,  {\em {Entanglement entropy of
  cosmological perturbations}}, Phys. Rev. D {\bf 102} (2020), no.~4, 043529
  [\href{http://www.arXiv.org/abs/2005.09688}{{\tt 2005.09688}}].

\bibitem{Mezei_2020}
M.~Mezei and G.~Sárosi,  {\em Chaos in the butterfly cone}, Journal of High
  Energy Physics {\bf 2020} (Jan, 2020).

\end{thebibliography}\endgroup

\end{document}